\documentclass[12pt, english, amsfonts]{article}

\usepackage[T1]{fontenc}
\usepackage{babel}

\def\H{{\cal H}}

\def\H{{\cal H}}

\def\ra{\rightarrow}

\def\d{\partial}

\def\b{\begin{eqnarray*}}  
\def\e{\end{eqnarray*}}    
\def\bn{\begin{eqnarray}}  
\def\en{\end{eqnarray}}   

\def\<{\langle}
\def\>{\rangle}

\def\no{\nonumber}

\def\v{\vskip1em}

\def\{{\lbrace}

\def\}{\rbrace}

\usepackage{graphicx}
\usepackage{rotating}
\begin{document}

\title{Proof that Half-Harmonic Oscillators become \\Full-Harmonic Oscillators 
after the Wall \\Slides Away}

\author{
  Carlos  R. Handy\\ 
  Department of Physics\\
  Texas Southern University \\
  3900 Cleburne St. \\
Houston, TX 77004 \bigskip
  \and
  John Klauder\\
  Department of Physics and Department of Mathematics  \\ 
University of Florida,   
Gainesville, FL 32611-8440}



\date{ }
\let\frak\cal

\maketitle 

\begin{abstract}
Normally, the half-harmonic oscillator is active when $x>0$ and absent when $x<0$. From a canonical quantization perspective, this leads to odd eigenfunctions being present while even eigenfunctions are absent. In that case, only the usual odd eigenfunctions will appear if the wall slides to negative infinity. However, if an affine quantization is used, sliding the wall away shows that all the odd and even eigenfunctions are encountered, exactly like any full-harmonic oscillator.
\end{abstract}.       
\newpage
\section{Implications of Canonical Quantization}\v
The full-harmonic oscillator has a classical Hamiltonian given by $H(p,q)=(p^2+q^2)/2$, where we have chosen the mass and frequency as $m=\omega=1$. Its familiar structure involves a complete set of  variables, i.e., $-\infty < p , q<\infty$, resulting in the standard, and well known, analysis.

The half-harmonic oscillator is different because, although manifestly identical, its variables are constrained to: 
$-\infty<p<\infty$ and $0< q<\infty$. In effect, this corresponds to imposing an infinite barrier potential along the nonpositive axis, $x \in (-\infty,0]$, leading to the constraint  $q=p=0$ for $x<0$.

When we implement a canonical quantization  on this problem, it follows that $Q>0$ results in a non-selfadjoint operator, $P \neq P^\dagger$, complicating the nature of the eigenvalue spectrum. Specifically,  two different Hamiltonian operators are possible, $\H_0=(P P^\dag + Q^2)/2$ and 
$\H_1=(P^\dag P +Q^2)/2$. Twice the spectrum for $\H_0$ is $2E_0=\hbar(1,5,9,...)$, while twice the spectrum for $\H_1$ is $2E_1 =\hbar(3,7.11....)$. Each of these two Hamiltonians leads to the same 
classical Hamiltonian, $H=(p^2+q^2)/2$, when $\hbar\ra 0$. Moreover, the spectrums can be intertwined such as $2E = \hbar(1, 5,9, 11,15,17,...)$, etc., which could lead to infinitely many spectrum versions! 

Clearly, canonical quantization, does not leat to a reasonable result for the half-harmonic oscillator. \v

\section{ Resolution through Affine Quantization}\v
In the context of the previous discussion, we see that although the momentum operator is not self-adjoint,  $P \neq P^\dag$, this difficulty can be easily corrected, by choosing a different operator. Our new operator is $D\equiv (P^\dag Q+QP)/2=(PQ+QP)/2$ because $P^\dag Q=PQ$, since  $P^\dag Q$ effectively fixes the fact that $P$ needs an operator that requires a null state.

The operator  $D\;(=D^\dag)$ can be seen from another point of view. The two operators from canonical quantization are $P$ and $Q$; and they obey the rule that $[Q,P]=i\hbar1\!\!1$. Upon multiplying this expression by $Q$ we obtain $i\hbar Q=(Q[Q,P]+[Q,P]Q)/2$, which becomes
$[Q,D]=i\hbar\,Q$. Unlike $Q$ for canonical quantization, which requires that $-\infty<Q<\infty$,
it follows that $[Q,D]=i\hbar Q$ requires that $Q>0$ or $Q<0$, or both. To stay valid the expression $[Q,P]=i\hbar 1\!\!1$
fails if multiplied by $Q=0$. 

Normally,  `$p$' and `$q$' are the classical variables promoted to canonical quantization. For affine quantization that privilege is given to `$pq$' and `$q$'; note well that `$p$', by itself, is {\it not} promoted in affine quantization.

Having $Q>0$ is just the rule that can help the half-harmonic oscillator.  This {\it positivity} criterion is a reccuring theme in the numerical analysis that follows.\v

\section{Affine Quantization and the \\Half-Harmonic Oscillator}\v
The Hamiltonian, once again, is $H(p,q)=(p^2+q^2)/2$, which, with $d\equiv pq$, now becomes $H'(d.q)= (d^2/q^2+q^2)/2$. The quantum version of the Hamiltonian is $\H=(DQ^{-2} D +Q^2)/2$.
Adopting Schr\"odinger's representation leads to $Q=x>0$ and 
 $D= -i\hbar [ x(\d/\d x)+(\d/\d x) x)]/2=-i\hbar [x (\d/\d x)+1/2]$. 
 
 The affine quantization of $\H$ becomes
    \bn && \H =\{ -\hbar^2 [x(\d/\d x)+1/2)] x^{-2} [x(\d/\d x)+1/2] + x^2\,\}/2 \no \\ \label{4}
           &&\hskip1.3em = \{ -\hbar^2 \d^2/\d x^2 + (3/4) \hbar^2/ x^2 + x^2 \}/2 \;,\en
  which is referred to as          
            a {\it spiked harmonic oscillator}, as recently studied by L. Gouba[1].
            She observed that this version of the half-harmonic oscillator 
             has eigenvalues of $2\hbar (n+1)$ for $ n =0,1,2,3,...$, which are equally spaced and 
             twice the spacing of the full-harmonic oscillator. Indeed, the $b = 0$ system corresponds to an {\it exactly solvable} system, and we can recover these exact energies without using the well known Nikiforov-Uvarov analysis [2] which relies on explicit wavefunction configuration representations. Instead, through our non-wavefunction based analysis, emphasizing certain algebraic properties of the power moments of the physical solutions, we can recover these exact eigenenergies as well.
             
             Our true interest is on a translated version  of Eq.(1) that allows the infinite barrier along $x<0$, to move  by an amount $b>0$, into the previously forbidden region ($x < 0$). That is, the new infinite potential barrier is imposed within the semi-infinite domain $x<-b$. This would transform equation Eq.(1) into the form:
               \bn \H = \{ -\hbar^2 \d^2/\d x^2 +(3/4)\hbar^2/(x+b)^2 +x^2 \}/2 \;, \en
with the discrete wavefunction boundary conditions: $\Psi(-b)  = 0$, as well as $\Psi(+\infty) = 0$.\v
      This more singular differential operator has not been solved in closed form (except for the $b = \{0,\infty\}$ case), to the best of our knowledge, so we have implemented a numerical study instead. Our question is: does increasing the value of $b$,  bring along both  odd and even full-harmonic oscillator expressions so that when $b\ra \infty$, the complete expression of the full-harmonic oscillator emerges, or not.  As detailed in the following algebraic and numerical study, we find that, indeed, in the asymptotic limit $b\rightarrow \infty$, the full set of even and odd harmonic oscillator solutions are recovered.\v
 Coincidentally, this numerical investigation will exploit a power moment representation of the system, focusing on important positivity (and nonnegativity) aspects of the system and its configuration space solutions. Additionally, the power moments define an affine map invariant representation space for the system, in keeping with the theoretical thrust of affine quantization. One important consequence, for the type of quantum potentials represented by Eq.(2) (i.e. differential systems with rational fraction coefficients, including extensions to multidimensions) is that we can generate tight lower and upper bounds on the discrete state energies; thereby insuring greater confidence in the nature of our results, particularly in the $b \rightarrow \infty$ limit, the regime of interest.

\section{Eigenenergy Bounding Formulations for ${\cal H}$ } \v

To facilitate the underlying numerical analysis, it will prove convenient to work with the coordinate variable $\chi \equiv x+b \geq 0$, resulting in the differential operator (i.e., $\hbar\rightarrow 1$)

             \bn \H = \{ -\d^2/\d \chi^2 +(3/4)\chi^2 +(\chi-b)^2 \}/2 \;, \en       
    for $b \geq 0$; with associated eigenenergy problem, ${\cal H}\Psi(\chi) = E \Psi(\chi)$, involving physical, $L^2$, exponentially bounded (decaying), wavefunctions satisfying  $\lim_{\chi\rightarrow \infty} \Psi(\chi) \rightarrow 0$, and $\Psi(0) = 0$. 

Due to the delicate question posed earlier concerning whether or not the even and odd harmonic oscillator states contribute in the $b \rightarrow \infty$ limit, we pursue a formalism that allows for the generation of tight bounds to the individual discrete states, in order to dispel any uncertainties. 

Two different bounding formalisms will be introduced. Each requires  that the differential system in question  (or a suitable transformation thereof, $\Psi \rightarrow F$) involve rational fraction coefficients (i.e., $\chi^{-2}$, $(\chi-b)^2$, in the present case). We will then focus on the power moments of the physical, configuration space, solution, 
\begin{eqnarray}
v(p) \equiv \int_0^\infty \ d\chi \ \chi^p F(\chi). 
\end{eqnarray}
For the types of systems being considered, these power moments will  satisfy a linear recursion relation of order $1+m_s$, involving known functions of the energy parameter, $E$:
\begin{eqnarray}
\sum_{\ell=0}^{m_s+1}v(p+\ell)\ \Omega_E(p,\ell) = 0, \ p \geq 0.
\end{eqnarray} 

This is referred to as the {\it moment equation representation} (MER) relation; and the starting point for either of the two bounding formalisms implemented in this work.

The focus on a MER based formalism is appropriate. In the present case, for $b = 0$, we know that the ground state solution is given by $E_{gr} = 2$
 and $\Psi_{gr}(\chi) = \chi^{3\over 2} exp(-{1\over 2 } \chi^2)$, up to a normalization, as confirmed through substitution.
 By working with the power moments of the contact transformation $F(\chi) \equiv  {\Psi}(\chi) {\Psi}_{gr}(\chi)$, the ensuing MER relation will quickly reveal the existence of the exact eigeneneriges, $E_n(b=0) =2(n+1)$, $n \geq 0$. This is discussed in Sec. (6.3.1).  The basic reason as to why this MER-$F$ representation can reveal the exact energies  is that most {\it exactly solvable} quantum systems involve discrete state solutions of the form ${\Psi}_n(\chi) = Q_n(\chi){\Psi}_{gr}(\chi)$ (i.e. $F_n(\chi) = Q_n(\chi)\Psi_{gr}^2(\chi)$), where the $Q_n$ are the orthonormal polynomials with respects to a weight corresponding to the square of the ground state configuration: $\langle Q_m|{\Psi}_{gr}^2|Q_n\rangle = \delta_{m,n}$. This structure is easily made manifest through the MER relation for the $F(\chi)$ configuration, and the important property of orthonormal polynomials: $\int_0^\infty d\chi \ \chi^m F_n(\chi) = 0$, for $m \leq n-1$, $n \geq 1$ [2]. 

The above is a general argument. In actuality, for the present system, the true form of the {\it exactly solvable} discrete states (for $b = 0$) will be ${\Psi}_n(\chi) = Q_n(\chi^2){\Psi}_{gr}(\chi)$, where the $Q_n$ are the $n$-th degree orthonormal polynomials  for a corresponding weight  defined on the  $\xi \equiv \chi^2 \geq 0$ domain.

We will expand upon the above, standard, but limited, use of orthonormal polynomial representations, through  the use of {\it weighted orthonormal polynomial expansions} of the form [3-6]
\begin{eqnarray}
\Psi(\chi) = \sum_{n=0}^\infty c_n P_n(\chi) R(\chi),
\end{eqnarray}
involving the orthonormal polynomials, $\{P_n\}$, relative to the weight $R$ (and not $R^2$, as is customary). This alternate approach will not only prove more flexible, but also allow for the exact generation of the expansion coefficients, $\{c_n\}$, through the underlying MER relation within the $\Psi$ representation. This in turn will allow us to generate tight lower and upper bounds to the various discrete state energies. This is the essence of our Orthonormal Polynomial Projection Quantization (OPPQ) analysis [4-6], described and applied, below. We stress that this OPPQ analysis comes in two basic forms. The first was developed by Handy and Vrinceanu [4,5] and defines an eigenenergy approximation method ansatz referred to here as OPPQ-AM. The more recent formalism, OPPQ-BM enables us to generate eigenenergy bounds, for arbitrary, multidimensional, discrete states, as developed by Handy [6].

As stated, two MER based, eigenenergy bounding formalisms are presented in this work. The first involves no explicit wavefunction basis expansion, and is referred to as the Eigenvalue Moment Method (EMM). It will yield important results that confirm the accuracy of the OPPQ formalism, which will allows for wavefunction reconstruction.

\subsection{The Eigenvalue Moment Method (EMM): Eigenenergy Bounds without a, Configuration Space, Basis Expansion}

It is possible to generate tight bounds for the discrete states, without the need for any basis expansion, by working within a power moment representation for a given system. The most direct implementation of this is the Eigenvalue Moment Method (EMM), as outlined below. It was developed by Handy and Bessis (HB) [7-10], based on an earlier discovery by Handy (i.e. ground state quantization through the Pade nesting property of Stieltjes measures),  and is considered to be one of the first applications  of semidefinite programming (SDP)  [11] analysis to quantum operators [12]; however, the underlying algorithms developed by HB were based on the use of linear programming (LP) [13], since robust SDP algorithms did not exist, as they do now.

We limit the following discussion to one dimensional Sturm-Liouville (SL) systems with rational fraction differential coefficients, or to those than can be transformed, $\Psi \rightarrow F$, into such form. The contact transformations of interest to us are of the form:

\begin{eqnarray}
F(\chi) = \cases{
(i)\  \ \ \Psi(\chi), \cr 
(ii)\  \ \Psi(\chi) {\cal R}(\chi), {\rm \ for \ some \ appropriate} \ {\cal R}(\chi) > 0,\cr 
(iii)\  \Psi^2(\chi),{\rm \ for \ selfadjoint \ SL\ systems},\cr
(iv)\  \ \Psi(\chi) \Psi^*(\chi), {\rm \ for \ non-selfadjoint \ SL \ systems.}\cr }
\end{eqnarray}

Define the the (physical) power moments as $\nu(p) \equiv \int_0^\infty \ d\chi \ \chi^p F(\chi)$. For the types of systems under consideration, either of these contact transformations will result in a moment equation recursion (MER) relation, of order $1+m_s$, representable in generator form
\begin{eqnarray}
\nu(p) = \sum_{\ell = 0}^{m_s} M_E(p,\ell) \ \nu_\ell.
\end{eqnarray}

It is important to guarantee that the chosen MER relation be unique for the physical states. This will usually be guaranteed by the fact that the power moments of unphysical solutions are infinite and thus cannot satisfy the MER relation; however, for some systems, such as the present one, there will be a subset of unphysical solutions for which almost all the power moments are finite and will satisfy the same physical MER relation, for almost all $p$-values. In such cases, it is important to initiate the MER recursion relation for $p$ - values unique to the physical states (i.e. $p \geq p_{initial}\equiv p_s$, where $|\nu_{physical}(p_s)| < \infty$ and $|\nu_{unphysical}(p_s)| = \infty$). 

An appropriate choice of normalization is also required (i.e. $\sum_{\ell = 0}^{m_s}\nu(\ell ) = 1$). This will then result in $m_s$ unconstrained missing moments. The final normalized MER relation takes on the form:
\begin{eqnarray}
\nu(p) = M_E(p,0)+\sum_{\ell = 1}^{m_s} {\hat M}_E(p,\ell) \ \nu_\ell,
\end{eqnarray}
corresponding, for instance, to $ {\hat M}_E(p,\ell) \equiv M_E(p,\ell)-M_E(p,0)$, $\ell \neq 0$.
The {\it initialization} moments, $\{\nu_\ell| 0 \leq \ell \leq m_s\}$ are referred to as the {\it missing moments}. The energy dependent coefficients, $M_E(p,\ell)$ are known in closed form. They will satisfy satisfy the MER relation as well, with regards to the `$p$' index, subject to the initialization conditions, $M_E(\ell_1,\ell_2) = \delta_{\ell_1,\ell_2}$.

 Depending on the chosen representation, $F(\chi)$, and the extent to which some, or all, of the discrete states correspond to nonnegative configurations, $F(\chi) \geq 0$, one can then impose positivity constraints arising from the {\it Moment Problem} (MP) in mathematics [14]; thereby generating tight constraints on the eigenenergies of those states with nonnegative configurations. 

Traditionally, the required  positivity constraints correspond to the nonlinear Hankel-Hadamard (HH) determinantal inequalities, of the general form:
\begin{eqnarray}
\Delta_{N;\sigma}(\nu) = Det\pmatrix{ \nu(\sigma)  & \nu(1+\sigma) & \ldots &\nu(N+\sigma) \cr \nu(1+\sigma)  & \nu(2+\sigma) & \ldots & \nu(N+\sigma+1) \cr  \cdot & \cdot & \ldots &  \cdot \cr \nu(N+\sigma)  & \nu(N+\sigma+1) & \ldots & \nu(2N+\sigma)} > 0,
\end{eqnarray}
$N \geq 0$, $\sigma = 0,1$. If the wavefunction is defined on the entire real axis, then only $\sigma = 0$ is required. For Stieltjes measures, corresponding to our system, then $\sigma = 0, 1$ are required.  Hausdorff problems (defined on a compact domain), require a third class of nonlinear  constraints. 

Upon substituting the MER relation into Eq.(10), as symbolized by the expression $\Delta_{N;\sigma}(E;\nu_ \ell) $, the ensuing nonlinear moment inequalities, constrain the physically allowed values for the energy and missing moments. That is, to given order $N$, for any energy parameter value, $E$, there will either exist, or not, missing moments satisfying the above inequalities. The missing moment solution set must be a nonlinear, bounded (through the normalization condition on the missing moments), convex set. We denote it by ${\cal V}_{N;E}$. The algorithmic objective is to determine the $E$-parameter values, at a given order $N$, for which the convex missing moment solution set exists (feasible), or does not exist. We then determine the endpoints of the feasible energy domain corresponding to the existence of such sets:
\begin{eqnarray}
{\cal E}_{N} = \{ E \ni  \Delta_{n;\sigma}(E;{\nu_\ell}) > 0| 0 \leq n \leq N, \sigma = 0,1\} .
\end{eqnarray}
The lower and upper bounds are generated from $E_N^{(L)} = Inf_E \Big( {\cal E}_{N}\Big)$, and  $E_N^{(U)} = Sup
_E \Big( {\cal E}_{N}\Big)$, resulting in
\begin{eqnarray}
 E_{N}^{(L)} < E_{N+1}^{(L)} \ldots < E_{phys} < \ldots < E_{N+1}^{(L)} < E_{N}^{(L)} .
\end{eqnarray}
the convergence is, generally, geometric with $N$. 

Each of the four $F$-formulations given in Eq.(7), where appropriate, combined with the HH constraints, corresponds to the Eigenvalue Moment Method.  The fourth contact transformation can be used on one dimensional non-selfadjoint differential operators. It was used to determine the first accurate prediction for the onset of PT symmetry breaking for  the potential $V(x) = ix^3+iax$  [15].

As an example, if one selects $F(\chi) \equiv \Psi(\chi)$, one can only generate tight bounds on the  ground state. This is because of the well known nodal structure of one dimensional Sturm-Liouville (SL) systems that insure that the ground state is positive, $\Psi_{gr}(\chi) > 0$, for quantum systems on the infinite real axis; and $\Psi_{gr}(\chi) \geq  0$, for systems on a semi-infinite domain (i.e. $\chi \geq 0$). 

One can also take $F(\chi) \equiv \Psi(\chi) {\cal R}(\chi)$, involving a well chosen, positive, function ${\cal R}(\chi) > 0$, usually emulating the asymptotic form of the physical ground state. This will generally produce faster converging bounds for the ground state, since the resulting {\it missing moment order}, $m_s$, is reduced. If $R (\chi) = \Psi_{gr}(\chi)$, and the quantum system is of the type {\it exactly solvable} or {\it quasi-exactly solvable}, the MER analysis for the corresponding $F$ representation will usually be able to identify the exact eigenergies of the system [16,17]. As noted, this is the case for the $b = 0$ parameter value for Eq.(3).

If the nodal structure of a particular discrete state is known, then one can choose an appropriate function, $R$, with signature, in order to generate a nonnegative configuration, $F \geq 0$, for the targeted discrete states. This can usually be done for parity invariant systems, when bounds for the first excited state are required.

If $F(\chi) \equiv \Psi^2(\chi)$, the corresponding, self-adjoint, SL problem will transform into a third order linear differential equation for the probability density. EMM can then be applied on the associated MER relation. Since each discrete state is now associated with  a nonnegative configuration, $F_{phys} \geq 0$, the corresponding eigenenergies can be bounded. 

We note that it is also possible to do the same for one dimensional, non-self adjoint, Sturm-Liouville problems. Through the Herglotz extension of the probability density, $F(\chi) = \Psi(\chi) \Psi^*(\chi)$, the corresponding $F(\chi)$ configuration will satisfy a fourth order linear differential equation, amenable to EMM.  This was used to correctly predict the onset of PT symmetry breaking for the quantum system $V(x) = ix^3+iax$ [15], as well as bound complex-valued Regge poles used in atomic scattering [17,18].

\section{Orthonormal Polynomial Projection Quantization (OPPQ): Eigenenergy Bounds and Wavefunction Reconstruction}

The previous EMM formulation can be extended to multidimensions, but only for bounding the bosonic ground state, since the corresponding wavefunction must also be nonnegative, $\Psi_{gr}({\overrightarrow r}) \geq 0$. This was used to generate the first accurate bounds to the ground state binding energy of hydrogenic atoms in superstrong magnetic fields (i.e. the quadratic Zeeman (QZM) effect). [9,10]

Since the multidimensional probability density,  $\Psi^2({\overrightarrow r})$, does not satisfy a linear partial differential equation, one cannot generate a MER representation, and thus EMM is non-implementable for multidimensional bosonic excited states, or fermionic states. Therefore, for a very long time, it was thought that the use of MER relations to bound appropriate multidimensional bosonic and fermionic excited states, was impossible. Very recently, it was realized that an alternate MER based analysis exists  that circumvents these EMM difficulties, as applied to the probability density configuration. This is referred to as the Orthonormal Polynomial Projection Quantization (OPPQ) Bounding Method (OPPQ-BM), and was used to study the multidimensional bosonic excited states of the QZM problem [6]. 

There are actually two versions of the basic OPPQ formulation: (i) one that is solely used for approximating the energies (i.e., OPPQ-Approximation Method, or OPPQ-AM); and (ii) another formulation that is used to bound the energies (i.e., OPPQ-BM). Each uses the same basis expansion in Eq.[6], as detailed below. However, each then implements the quantization process through related, but very different, relations. OPPQ-BM can also be used to approximate the energies. All of these OPPQ formulations are considered here, as applied to the system in Eq.(3). 

The OPPQ analysis can be applied to either the $F(\chi) = \Psi(\chi)$ representation, or the $ F(\chi)=  \Psi(\chi){\cal R}(\chi)$ representation (i.e. case (ii) in Eq.(7)). In this work, we will implement OPPQ on the first, since it is more directly connected to the actual wavefunction configuration of interest.

 We now reconsider Eq.[6] corresponding to a complete [20], non-othonormal (weighted-polynomial) basis expansion, the foundation for Orthonormal Polynomial Projection Quantization : 

\begin{eqnarray}
\Psi(\chi) = \sum_{j=0}^\infty c_n {\cal B}_n(\chi),
\end{eqnarray}
where the basis functions are define by
\begin{eqnarray}
{\cal B}_n(\chi) \equiv P_n(\chi) R(\chi),
\end{eqnarray}
involving the orthonormal polynomials
\begin{eqnarray}
P_n(\chi) = \sum_{j=0}^n \Xi_j^{(n)} \chi^j,
\end{eqnarray}
with respect to an arbitrarily chosen, nonnegative, exponentially decaying, weight, $R(\chi) \geq 0$.
The polynomial orthonormality conditions  correspond to: 
\begin{eqnarray}
\langle P_m|R|P_n\rangle = \delta_{m,n}.
\end{eqnarray}
We emphasize that $R(\chi)$ is arbitrary, although  faster converging results ensue if it is  chosen to emulate the asymptotic behavior of the physical state(s) of interest:
\begin{eqnarray}
R(\chi) \sim \Psi_{phys}(\chi) .
\end{eqnarray}
It bears repeating that Eq.(13) does not involve an expansion in terms of an orthnormal basis. Indeed, the chosen basis ${\cal B}_n(\chi) \equiv P_n(\chi) R(\chi)$ is non-orthonormal, due to the condition in Eq.(16). That is:
\begin{equation}
\langle {\cal B}_{m}|{\cal B}_n\rangle \neq \delta_{m,n}.
\end{equation}
This then allows us to generate the expansion coefficients in Eq.(13) {\it exactly} as functions of the energy parameter and the missing moments:

\begin{eqnarray}
c_n(E;\nu_\ell) & = & \int_0^\infty d\chi  \ P_n(\chi) \ \Psi(\chi), \\
 & = & \sum_{j=0}^n \Xi_j^{(n)} \nu(j), \nonumber \\
& = & \sum_{j=0}^n \Xi_j^{(n)} \Big(\sum_{\ell = 0}^{m_s} M_E(j,\ell)\  \nu_\ell \Big),  \\
&  = & \sum_{\ell=0}^{m_s} \Lambda_\ell^{(n)}(E) \ \nu_\ell, \nonumber \\
 & = & {\overrightarrow \Lambda}^{(n)}(E) \cdot {\overrightarrow \nu}, 
\end{eqnarray}
where
\begin{eqnarray}
\Lambda_\ell^{(n)}(E) = \sum_{j=0}^n \Xi_j^{(n)}M_E(j,\ell),
\end{eqnarray}
and
\begin{eqnarray}
{\overrightarrow \nu} = (\nu_0,\nu_1, \ldots, \nu_{m_s}).
\end{eqnarray}
We note that Eq.(20) makes use of the MER representation. We emphasize that Eq.(21) is the exact expression for the OPPQ projection coefficients as functions of the energy parameter (involving known coefficients) and the missing moments.

Quantization ensues from demanding that the weight be chosen so that the integral 
\begin{eqnarray}
{\cal I}[\Psi, R] \equiv \int_0^\infty d\chi\  {{\Psi^2(\chi)}\over {R(\chi)}} = \cases{ finite \iff  \ \Psi = \Psi_{phys} ,\cr
\infty \iff \ \Psi \neq  \Psi_{phys} ,\cr}
\end{eqnarray}
be finite only for the physical solutions (i.e. energy and missing moment values); and infinite, if not. 

More generally, if $R(\chi)$ decreases (essentially) no faster than the asymptotic form of the physical solutions, 

\begin{eqnarray}
Lim_{\chi \rightarrow \infty} {{|\Psi_{phys}(\chi)|}\over{R(\chi)} }\rightarrow const,
\end{eqnarray}
with $0 \leq const < \infty$,
then Eq.(24) can discriminate between the physical and unphysical solutions.
The preferred choice, leading to faster convergence, is to choose the weight to emulate the asymptotic form of the physical configuration (i.e., $const \neq 0$) [4,5]. 

If we substitute Eq.(13) in the integrand in Eq. (24) and make use of the orthonormal polynomial property in Eq.(16), the following positive series representation for the integral follows:

\begin{eqnarray}
 {\cal I}[\Psi, R]  & = & \sum_{n=0}^\infty \ c_n^2 [E,\nu_0,\ldots,\nu_{m_s}], \\
   & = & \cases{finite, \iff \ E= E_{phys}\ and \ \nu_\ell = \nu_{\ell,phys}, \cr
			\infty, \iff \ E \neq E_{phys}\ or \ \nu_\ell \neq \nu_{\ell,phys} \cr}.
\end{eqnarray}

Both Eqs.(26,27) define, effectively, a {\it shooting method} ansatz in which for arbitrary selection of $E$ and $\{\nu_0,\ldots,\nu_{m_s}\}$, one generates the  positive partial sums in Eq.(26) and determines if they converge or not. Clearly, this is only suggestive, since the {\it missing moment} order can be large, even for one dimensional systems. A more efficient analysis is required that elliminates having to test Eq.(27) over all of the missing moment space and energy paramater space as well. This is the major focus of the following discussion.

\subsection{The OPPQ-Approximation Method (OPPQ-AM):}
Clearly, a different interpretation of Eq.(27) is that for the physical solutions we must have

\begin{eqnarray}
Lim_{n\rightarrow \infty} c_n(E,{\overrightarrow \nu}) = 0.
\end{eqnarray}
Given Eq.(21) this leads to the determinantal condition:
\begin{eqnarray}
Lim_{n\rightarrow \infty} Det\Big( \Lambda^{(n-\ell_1)}_{\ell_2}(E_{phys}) \Big) = 0,
\end{eqnarray}
involving the $(1+m_s) \times (1+m_s)$, energy dependent matrix, $\Lambda^{(n-\ell_1)}_{\ell_2}(E)$, for sufficiently large `$n$' values. The approximate energies derived in this manner constitute the Orthonormal Polynomial Projection Quantization - Approximation Method (OPPQ-AM) [4,5]. 

\subsection{The OPPQ-Bounding Method (OPPQ-BM):}

A different quantization procedue also ensues from Eq.(27), leading to the generation of converging bounds to the various discrete states: OPPQ-BM.  However, as noted, OPPQ-BM also contains an alternate eigenenergy approximation formalism, to be referred to as the OPPQ-BM/Approimation Method. The significance of these two approximation methods  is that although the results of OPPQ-AM (i.e. Eq.(29)) tend to converge fast (and more so, for weights, $R$, that emulate the physical asymptotic form), they do not guarantee that the approximate eigenenergies generated are purely real. In the asymptotic limit, any spurious imaginary parts disappear. The alternate approximation method, that arising within OPPQ-BM, as defined below, will always generate real eigenenergy approximants. Having said this, the OPPQ-AM estimates are easier to generate and appear to be faster converging than the  second approach; although the second is more rigorous, since it only produces real eigenenergies.

The starting point for the OPPQ-BM formalism is to focus on the positive, partial sum, sequence elements

\begin{eqnarray}
S_I(E,{\overrightarrow \nu}) \equiv \sum_{n=0}^I c_n^2 = \langle{\overrightarrow \nu} | {\cal P}^{(I)}(E)|{\overrightarrow \nu}\rangle,
\end{eqnarray}
where  the positive definite matrix is the sum of the individual dyads:

\begin{eqnarray}
{\cal P}^{(I)}(E) = \sum_{n=0}^I {\overrightarrow \Lambda}^{(n)}(E) {\overrightarrow \Lambda}^{(n)}(E) .
\end{eqnarray}
These partial sum sequences are naturally non-decreasing:
\begin{eqnarray}
S_{I}(E, {\overrightarrow \nu}) < S_{I+1}(E, {\overrightarrow \nu}) < \ldots.
\end{eqnarray}
They will only be bounded from above if the energy and missing moments correspond to the true physical values, according to Eq.(27). We can determine the physical energies through a more efficient process that, essentially, factors out the missing moment dependence.

If we adopt a unit missing moment vector normalization $|{\overrightarrow \nu}|^2 = 1$, then we can restrict the analysis to the smallest eigenvalue of the above positive definite matrix:

\begin{eqnarray}
\lambda_I(E) = \ Smallest\ Eigenvalue \ of \ {\cal P}^{(I)}(E).
\end{eqnarray}
These expressions also form a positively increasing sequence, for any $E$:
\begin{eqnarray}
0 < \lambda_I(E) < \lambda_{I+1}(E) < \lambda_{I+2}(E) < \ldots \lambda_\infty(E).
\end{eqnarray}
Thus, the eigenvalue functions, $\lambda_I(E)$ , form a nested, concaved upwards (i.e. relative to the local minima), sequence of functions that become unbounded everywhere except at the physical energies. This is depicted in Fig.  2 and Fig. 3, as discussed in Sec. 8.4.

It then follows that Eqs.(26,27) transform into the equivalent expression:
\begin{eqnarray}
Lim_{I\rightarrow \infty}\lambda_I(E) = \cases{ finite, \iff E = E_{phys} \cr
\infty, \iff E \neq E_{phys} \cr}.
\end{eqnarray}
All that is required is to search within the energy parameter space for those energy values which satisfy Eq.(35); however, we can identify yet two other, better procedures,  for achieving this. One leads to a second eigenenergy approximation method; while the other results in an eigenenergy bounding ansatz.

\subsubsection{Using OPPQ-BM to Approximate the Eigenenergies}
Given Eq.(34) and Eq.(35), the natural focus is on the minimia extremum values:
\begin{eqnarray}
\partial_E\lambda_I(E_I^{(min)}) = 0,
\end{eqnarray}
which will define the $I$-th order approximant to the discrete state energy. This is a different approximate quantization condition to the OPPQ-AM procedure in Eq.(29). It will be referred to as the {\it OPPQ-BM-Approximation Method}. Clearly, the eigenenergy approximants to Eq.(36) will always be real, as opposed to the possible generation of spurious imaginary parts to the energy approximants derived from Eq.(29) (although such spurious imaginary parts are expected to disappear in the asymptotic limit, $n \rightarrow \infty$).

\subsubsection{Using OPPQ-BM to Bound the Eigenenergies}
The $\lambda_I(E_I^{min})$ also satisfy a specific, non-decreasing, positive, and {\it bounded from above} sequence:

\begin{eqnarray}
\lambda_I(E_I^{(min}) < \lambda_{I+1}(E_{I+1}^{(min})  < \ldots < \lambda_\infty(E_{phys}) < \infty.
\end{eqnarray}
One can use Eq.(37) to generate empirically determined converging bounds on the true physical energy. 

Let ${\cal B}_U$ be any, empirically determined, coarse upper bound, to the  convergent expression (i.e. for a chosen discrete state):
\begin{eqnarray}
\lambda_I(E_I^{(min}) < \lambda_{I+1}(E_{I+1}^{(min})  < \ldots < \lambda_\infty(E_{phys}) < {\cal B}_U.
\end{eqnarray}
Given Eq.(35), there will always be $E$ values satisfying:
\begin{eqnarray}
\lambda_I(E_I^{(L)}) = \lambda_I(E_I^{(U}) = {\cal B}_U.
\end{eqnarray}
These will then correspond to converging lower and upper bounds to the physical energy:
\begin{eqnarray}
E_I^{(L)} < E_{phys} < E_I^{(U)},
\end{eqnarray}
with
\begin{eqnarray}
Lim_{I \rightarrow \infty} E_I^{(U)}-E_I^{(L)} = 0^+.
\end{eqnarray}
\newpage

\section {The Eigenvalue Moment Method (EMM): Numerical Results}

In this section we implement EMM on three configuration space representations: $F(\chi) \equiv \{ \Psi(\chi),  R(\chi)\Psi(\chi), \Psi^2(\chi)\}$, as outlined earlier. 

The first, $F = \Psi$, will only generate tight bounds for the ground state, for various ranges of the `$b$' parameter. 

The second contact transformation (for different choices of $R > 0$) will generate improved (tighter) bounds for the ground state.  In particular, when $R(\chi) \equiv \Psi_{gr}(\chi)$ we will be ale to confirm the {\it exactly solvable} nature of the underlying system for $b = 0$.

The third formulation, $F = \Psi^2$, will generate bounds for arbitrary excited states as well. 

The above  results will prove valuable in better assessing the accuracy, efficiency, and effectiveness of the subsequent OPPQ analysis. 

The EMM approach does not require an explicit basis expansion. The basic EMM philosophy is to combine the underlying  MER relation for the physical power moments, with relevant positivity criteria  (i.e. the {\it Moment Problem},  Positivity, HH conditions) unique to quantizing the some, or all, physical solutions. 

The same philosophy applies within OPPQ analysis. It will use the MER relation, combined with the finiteness of a certain positivity condition (i.e. Eq.(24)) unique to the physical states. In practice, EMM appears to be faster converging than OPPQ; however, the extension of EMM to $\Psi^2$ is not as robust as OPPQ. All of this will become apparent in this section, dedicated to EMM, and the following section which implements OPPQ.

\subsection{EMM-$\Psi(\chi)$}
The translated eigen-system, $\chi \equiv  x+b \geq 0$, for the hamiltionian in Eq.(3) corresponds to
\begin{equation}
- \partial_\chi^2\Psi + \Big({3\over 4}{1\over{\chi^2}} + (\chi^2+\beta \chi)\Big)\Psi= \lambda \Psi,
\end{equation}
where $\beta \equiv -2 b$ and $\lambda \equiv 2E -b^2$. The origin corresponds to a regular singular point, generating independent solutions of the form
\begin{eqnarray}
Y_j(\chi) = \cases {\chi^{3\over 2} A_1(\chi), \ for \ j = 1 , \cr
\chi^{3\over 2} Ln (\chi) A_1(\chi) + \chi^{-{1\over 2}} A_2(\chi), \ for \ j = 2 ,}
\end{eqnarray}
where the $A_j$ are analytic near the origin. The power series expansion for $A_1$ is uniquely determined, and it in turn generates $A_2$. 

Our immediate objective is to generate a moment equation for the physical solutions to Eq.(42). 
Any solution to Eq.(42) must be a linear superposition of these two independent configurations, $\Psi(\chi) =c_1 Y_1(\chi)+c_2 Y_2(\chi)$. The physical, $L^2$ solutions, must asymptotically vanish at infinity according to the zeroth order WKB expression $\Psi(\chi) \rightarrow exp(-{1\over 2}\chi^2)$. The $L^2$ condition, $\int_0^\infty \Psi^2(\chi) < \infty$, leads to the following conditions for the physical solutions:

\begin{eqnarray}
\Psi_{phys}(\chi) \sim \cases{ \chi^{3\over 2} A_1(\chi),\  \chi \rightarrow 0^+ , \cr
{\cal N} exp(-{1\over 2}\chi^2), \ \chi \rightarrow \infty}.\cr
\end{eqnarray}
Despite this, there will be  physical and unphysical solutions of exponentially decaying,  asymptotic form (i.e. $exp(-{{\chi^2}\over 2})$)
, with (almost all) finite power moments. This is in contrast to those unphysical solutions that are exponentially unbounded in the asymptotic limit.

 Define the power moments as:

\begin{eqnarray}
v(p) \equiv \int_0^\infty d\chi \ \chi^p \Psi(\chi).
\end{eqnarray}
Based on the local analysis in Eq.(43), the power moments for the aforementioned class of (asymptotically, exponentially, bounded) unphysical solutions,  will be finite so long as the combined singularity of $\chi^p\Psi(\chi)$, at $\chi = 0$, is {\it soft}; or alternatively:
\begin{eqnarray}
 |v_{unphys}(p)| < \infty, \ if \ p > -{1\over 2}.\cr
\end{eqnarray}
The power moments of the physical solutions will be finite if an analogous, {\it soft} singularity condition is satisfied:

\begin{eqnarray}
|v_{phys}(p)| < \infty, \ if \ p > -{5\over 2} .
\end{eqnarray}

Define the power moments on the semi-bounded domain $\Re_\delta \equiv [\delta,\infty)$:
\begin{eqnarray}
v_\delta(p) \equiv \int_\delta^\infty d\chi \ \chi^p \Psi(\chi). 
\end{eqnarray}
Multiplying Eq.(42) by $\chi^{p}$ where $p \in (-\infty,+\infty)$, and integrating by parts over the $\Re_{\delta}$ domain (i.e. assuming $\Psi \sim exp(-\chi^2/2)$),  generates the following moment equation relation (MER) 
\begin{eqnarray}
 \hspace{-20pt} \delta^{p}\Psi'(\delta)-p \delta^{p-1}\Psi(\delta) +\Big(3/4 -p(p-1)\Big) v_\delta(p-2) +v_\delta(p+2) \\ \nonumber
\hspace{20pt}+\beta v_\delta(p+1)  = \lambda v_\delta(p) ,\cr
\end{eqnarray}
for $-\infty < p < + \infty$. We note the boundary terms for the physical, $Y_1(\delta) \sim \delta^{3\over 2}A_1(\delta)$, and the unphysical,  $Y_2(\delta) \sim \delta^{3\over 2}Ln(\delta) A_1(\delta)+ \delta^{-{1\over2}}A_2(\delta)$, solutions. 

Through a straightforward analysis we can conclude that the physical solutions will satisfy the $\delta \rightarrow 0^+$ moment equation relation (MER):
\begin{eqnarray}
 v(p+2)=   -\beta v(p+1)   + \lambda v(p) + \Big(p(p-1) -{3\over 4}  \Big)v(p-2),
\end{eqnarray}
$p > -1/2$. 

It is important to note that the same MER relation is satisfied by the unphysical, asymptotically exponentially bounded solutions, if $p > 3/2$. That is, their power moments will also be finite, and satisfy the same MER, but only starting at $p > 3/2$. The physical solutions satisfy this same MER but starting at a lower initial $p$-value, $p > -1/2$. In fact, this MER relation for the physical solutions must initiate for $p \geq p_{initial} \equiv p_s $ where $-1/2 <  p_s \leq 3/2$, otherwise it will not filter out the unphysical solutions (which will not satisfy Eq.(50) for this limited range of $p$-values).  In particular, $p_{s} = \{0,{1\over 2}\}$ are possible initial values (i.e. the integer, as well as the half-integer, order power moments couple amongst themselves through their respective MER relations). 

The fact that both physical and unphysical solutions satisfy the same MER relation, but initiated at different $p$-values,  suggests that this type of analysis requires high accuracy computation, since the asymptotic power moments, $v(p \rightarrow \infty)$ may become less sensitive to the $\{v(p_s-2),v(p_s-1)
\}$ contributions, if insufficient accuracy in the calculations are ignored. 

The EMM formalism requires that we work with the nonnegative integer power moments of an appropriate configuration. If $p_s = 0$, Eq.(50) would require working with the $v(-1)$ and $v(-2)$ power moments. To avoid modifying the EMM formalism (i.e. to accomodate negative integer order power moments), we simply work with a slightly modified configuration. To this extent, define alternative power moments for the physical solutions: 
\begin{eqnarray}
u(p) \equiv  \int_0^\infty d\chi \chi^{p} \chi^{-2}\Psi_{phys}(\chi),
\end{eqnarray}
for $p \geq 0$. If we take $v(p) \equiv u(p+2) $, then the corresponding $u$-moment equation becomes:
\begin{eqnarray}
 u(p+4)=   -\beta u(p+3)   + \lambda u(p+2) + \Big(p(p-1) -{3\over 4}  \Big)u(p),\cr
\end{eqnarray}
$p > -{1\over 2}$, although we will restrict $p = integer \geq 0$. 

\subsubsection{ EMM-$\Psi$, $b = 0$}
For $b = 0$, Eq.(52) becomes:

\begin{eqnarray}
 u(p+4)=      2E u(p+2) + \Big(p(p-1) -{3\over 4}  \Big)u(p),\cr
\end{eqnarray}
$p > -{1\over 2}$. 

If we restrict  $p = integer \geq 0$, then Eq.(53) separates into two recursion relations, one for the even order (Stieltjes) power moments, and another for the odd order power moments. Each is of order 2, since in the first case $\{u(0),u(2)\}$ must be specified in order to generate all the other even order moments; whereas  the odd order moments $\{u(1),u(3)\}$ must be specified, before the other odd order moments can be generated. 

Let us denote the even and odd order moments by $h_\sigma(q) = u(2q+\sigma) $, where $\sigma =\{0,1\}$, we obtain:

\begin{eqnarray}
 h_\sigma(q+2)=    2E \ h_{\sigma}(q+1) + \Big((2q+\sigma)(2q+\sigma-1) -{3\over 4}  \Big)h_\sigma(q),\cr
\end{eqnarray}
$q \geq 0$.

\subsubsection{ EMM-$\Psi$, $b \neq 0$}
For $b \neq 0$ and $p = integer \geq 0 $, Eq.(52) corresponds to a finite difference equation of order 4, since the initialization moments $\{u(\ell)|0 \leq \ell \leq m_s \equiv 3\}$ must be specified before all the other moments can be generated. As previously noted, we refer to the initialization moments as the {\it missing moments}, $u_\ell \equiv u(\ell)$. We can generate the power moments through the relation
\begin{eqnarray}
u(p) = \sum_{\ell =0}^{m_s = 3} M_\lambda(p,\ell) \ u_\ell,
\end{eqnarray}
where the $\lambda = 2E-b^2$ dependent coefficients, $M_\lambda(p,\ell)$ satisfy Eq.(52), with respects to the $p$-index,  subject to the initialization conditions, $M_\lambda(\ell_1,\ell_2) =\delta_{\ell_1,\ell_2}$, for $0 \leq \ell_{1,2} \leq 3$. 

A normalization condition needs to be imposed. Although the physical choice in configuration space is $\langle \Psi|\Psi \rangle =1$, working in a moments' representation requires an alternate choice. Since the ground state is a nonnegative configuration, one suitable choice of normalization is:

\begin{equation}
\sum_{\ell = 0}^{m_s =3} u_\ell = 1.
\end{equation}

The results of implementing EMM on the $\Psi$ representation is given in Table 1, for various values for $b= 0, 10, 20,  \ldots$. The tightness of the eigenenergy bounds for the ground state are limited by the low machine precision (i.e. 14 digits) of our Fortran code, and low moment expansion order indicated $P_{max} < O(30)$. Nevertheless, these results provide a useful guide.

\begin{table}
\caption{
$EMM-\Psi_{gr}$.
}
\centerline{
\begin{tabular}{rrrrr}
\hline
$b$ & $E^{(L)}$    &  $E^{(U)}$  &   $P_{max}$  & Equation \\
\hline
 0& 1.999415&  2.000489&  29 & [52]\\
0 &  1.882144&2.065301 & 35&  $\sigma = 0$ [54]\\
0 &  1.946320& 2.199344 & 36&  $\sigma = 1$ [54]\\
\hline
.1 & 1.870371 & 1.871507& 28 & [52]\\
.5 & 1.428965 & 1.429646 & 30 & [52]\\
1 & 1.032844 & 1.033323 & 28 & [52]\\
5 & .515648 & .516333 & 23 & [52]\\
 10& .496295 & .524709 & 16& [52] \\
20 & .457391&  .658807 & 13& [52] \\
 \hline
\hline
\end{tabular}}
\end{table}

\subsection{EMM for the $F(x) \equiv \Phi(x) = exp(-{{x^2}\over 2}) \Psi(x) $ Representation}

An alternative EMM strategy for improving the tightness of the bounds is to choose a representation consistent with the EMM theory but involving fewer missing moments. One way to do this is to work with a contact transformation incorporating the asymptotic form (at infinity) of the physical solutions: $\Phi(x) = exp(-{{x^2}\over 2}) \Psi(x) $. This will generally result in a MER relation with reduced missing moment order.

In choosing the appropriate contact transformation, one must insure that physical states will retain the finiteness of the power moments; whereas unphysical states will retain the infiniteness of all, or some, of their power moments. This is generally satisfied by taking $R(x) \sim \Psi_{phys}(x)$. As a counterexample, for the system in question, choosing a weight of $exp(-x^4)$ would violate this condition, since all the unphysical solutions (including the unbounded solutions) in the $\Psi$ representation would have finite power moments in the transformed representation; although the physical states would retain the finiteness of their power moments.

We recall that for the $b = 0$ case ($\chi = x$), we know that the exact ground state wavefunction, up to a normalization, is $\Psi_{gr}(\chi) = \chi^{3\over 2} exp(-{1\over 2} \chi^2)$. 

If  one wants to study if the discrete states correspond to that of an {\it exactly solvable} system, one should use a contact transformation involving the actual ground state (nonnegative) wavefunction:  ${\tilde \Phi}(x) \equiv \Psi_{gr}(x)\Psi(x) =x^{3\over 2} exp(-{{x^2}\over 2}) \Psi(x)$. This is because for {\it exactly solvable} systems, the discrete states are usually of the form $\Psi_n(x) = Q_n(x) \Psi_{gr}(x)$, involving polynomial factors, $Q_n$, corresponding to some, or all, of the orthonormal polynomials of the weight $R^2 \equiv \Psi_{gr}^2$  (i.e. $\langle Q_m|R^2|Q_n\rangle = \delta_{m,n}$). Studying the MER relation for ${\tilde \Phi}$ will then reveal the hidden {\it exactly solvable} nature of the system. This is derived below.

The following analysis exploits the $\Phi$ representation, as defined above. However, the ensuing moment relations also pertain to those of ${\tilde \Phi}$ through a simple translation: ${\tilde v}(p) = v(p+{3 \over 2})$, for $ p \geq 0$.

\subsection{The MER Formulation for $\Phi(x) = exp(-{{x^2}\over 2})\Psi(x)$}

Our original problem was 
\begin{equation}
- \partial_x^2\Psi + \Big({3\over 4}{1\over{(x+b)^2}} + x^2\Big) \Psi= 2E \Psi,
\end{equation}
for $x+b \geq 0$, with $\Psi(-b) = 0$.  Let us work with $\Phi(x) \equiv exp(-{{x^2}\over 2}) \Psi(x)$, over the same domain, and then do a change of variables into $\chi = x+b$, for $x \geq -b$. The contact transformation is modeled after the asymptotic form of the wavefunction $\Psi(x) \sim  exp(-{{x^2}\over 2})$.

We can transform the Schrodinger equation into a differential equation for $\Phi(\chi) = exp\big( -{1\over 2}(\chi-b)^2\big)\Psi(\chi)$, $\chi \equiv x+b \geq 0$:

\begin{eqnarray}
-\Big(\Phi''(\chi) + 2 (\chi-b)\Phi'(\chi)\Big) + {3\over 4}{1\over{\chi^2}}\Phi(\chi) = (2E+1) \Phi(\chi),
\end{eqnarray}
$\chi \geq 0$.
Since the $\Phi$ representation differs from the $\Psi$ representation by an analytic factor, the local structure (i.e. $\chi \approx 0$) of the $\Phi$ solutions is identical to that of the $\Psi$'s. 
Implementing the same analysis as before, upon multiplying both sides by $\chi^{p}$ and integrating by parts over the $\Re_{\delta}$ domain yields the moment equation 

\begin{eqnarray}
 \delta^{p}\Phi'(\delta)+\Big(2\delta^{p+1}-2b\delta^{p}-p \delta^{p-1}\Big)\Phi(\delta)  -2bpv_\delta(p-1)  \\ \nonumber
+\Big({3\over 4}-p(p-1)\Big) v_\delta(p-2)  = (2E-1-2p) v_\delta(p) ,\cr
\end{eqnarray}
where $v_\delta(p) =\int_\delta^\infty d\chi \ \chi^p\Phi(\chi)$. 

Applying a similar analysis to that for the  EMM-$\Psi$ representation, discussed in the previous subsection, leads us to conclude that the appropriate MER relation for the physical solutions  corresponds to (i.e. $v(p) \equiv lim_{\delta \rightarrow 0^+}v_\delta(p)$)

\begin{eqnarray}
\hspace{-30pt} (2E-1-2p) v(p) = -(p+1/2)(p-3/2)  v(p-2)-2bpv(p-1),
\end{eqnarray}
for $p > -1/2$.  As before, we require $p \geq p_{initial} \equiv p_s$ where $-1/2 < p_{s} \leq 3/2$, since initiating the recursion relation at higher values would not filter out the unphysical solutions that are exponentially decaying at infinity. Only moment recursion relations unique to the physical solutions are allowed within EMM. 

The MER in Eq.(60) is valid for all discrete states. The choice of initial $p$-index value, $p_s > -1/2$, determines the set of power moments to be used in describing the discrete state of energy $E$. Thus, different $p_s$ values will generate different sets of power moments with which to describe the same energy state. The question then becomes, which of these is the most efficient or effective.

If $b \neq 0$, then the power moments $v(p_s-2), v(p_s-1), v_(p_s), \ldots$, couple amonst themselves. That is, $\nu_{p_s}(n) \equiv v(n+p_s) = \int_0^\infty d\chi  \ \chi^n  \times \chi^{p_s} \Phi(\chi)$, for $n = -2,-1,0,\ldots$, all couple amongst themselves, and satisfy:

\begin{eqnarray}
 (2E-1-2(n+p_s)) v_{p_s}(n) = -(n+p_s+1/2)(n+p_s-3/2)  v_{p_s}(n-2) \nonumber \\  
-2b(n+p_s)v_{p_s}(n-1),\cr
\end{eqnarray}
for $n \geq 0$. 

 As indicated above, $-1/2 < p_s \leq 3/2$, is the required condition. In particular, $p_s = 0$ and $p_s = 1/2$ define two distinct groups of power moments (i.e. the integer and half-integer order groups) corresponding to the same physical, discrete, state.

\subsubsection {Proof that $b = 0$ corresponds to an {\it Exactly Solvable} quantum system }

When $b = 0$ these power moments satisfy the MER relation

\begin{eqnarray}
 (2E-1-2(n+p_s)) v_{p_s}(n) = -(n+p_s+1/2)(n+p_s-3/2)  v_{p_s}(n-2),\cr
\end{eqnarray}
for $n \geq 0$; and the moments  $\{v_{p_s}(2n)| n = -1,0,1,\ldots\}$ will couple amongst themselves. Three different choices emerge, consistent with the requirement  $-1/2 < p_s \leq 3/2$: $p_s \in \{ 0, {1\over 2},1\}$.

If we take $p_s = 3/2$ and $n = 0$ in Eq.(62),  we obtain 
\begin{eqnarray}
(E-2) v_{3\over 2}(0) = 0.
\end{eqnarray}
The nonnegativity of the ground state wavefunction, and the resulting positivity for all its power moments, requires $v_{1\over 2}(0) > 0$. Therefore, the ground state energy must correspond to  $E_{gr}(b=0) = 2$; whereas all the excited states must satisfy 
\begin{eqnarray}
v_{3/2}^{(exc)}(0) \equiv {\tilde v}_{exc}(0) = 0,
\end{eqnarray}
 the zeroth moment of the configuration ${\tilde \Phi}(x) = \Psi_{gr}(x) \Psi(x)$, as defined at the outset.

We can extend the previous eigenenergy argument to the $b = 0$ excited states as follows. Clearly, Eq.(62) couples the even or odd order moments among themselves. For the even order power moments, $n =2\eta$, and $p_s = 3/2$, we obtain
\begin{eqnarray}
 (E-2(\eta+1)) v_{3/2}(2\eta) = -\eta(2\eta+2)  v_{3/2}(2(\eta-1)),\cr
\end{eqnarray}
for $\eta \geq 0$. 

Since any excited state must satisfy $v_{3/2}^{(exc)}(0) = 0$, unless the excited state energy, $E_{\eta \neq 0}$, satisfies the zero coefficient condition, $E_\eta-(2\eta+1) = 0$, for some $\eta = n \geq 1$, all the power moments will be zero. This would then lead to the conclusion that the wavefunction must be zero, which is impossible for the $L^2$ physical solutions. Therefore, we conclude that the discrete state energies must be given by the exact relation:

\begin{eqnarray}
E_n = 2(n+1),
\end{eqnarray}
for $n \geq 0$ and $E_0 \equiv E_{gr}$.

With regards to the form of the wavefunction, we
note that the even order moments $v_{3\over 2}(2n) $ are the ordinary Stieltjes power moments of the corresponding configuration on the $\xi \equiv x^2 \geq 0$, domain (i.e. $ b = 0$):

\begin{eqnarray}
v_{3\over 2}(2\eta) \equiv \mu_\eta =\int_0^\infty d\chi \chi^{2\eta + {3\over 2}} \Phi(\chi)  \equiv \int_0^\infty d\xi \ \xi^{\eta}\ \Upsilon (\xi),
\end{eqnarray}
$\Upsilon(\xi) ={1\over 2} \xi^{1\over 4} \Phi(\xi) = {1\over 2} \xi^{-{1\over 2}} \Psi(\xi)\Psi_{gr}(\xi)$. The  ground state corresponds to $\Upsilon_{gr}(\xi) ={1\over 2} {1\over {\sqrt{\xi}}} \Psi_{gr}^2(\xi)$. The $\mu_\eta$ moments will now satisfy

\begin{eqnarray}
(E-2(\eta+1)) \mu_\eta = -2\eta(\eta+1) \mu_{\eta-1},
\end{eqnarray}
for $ \eta \geq 0$. However, the analysis for Eq.(65) results in the solutions:
\begin{eqnarray}
\mu_\eta^{(n)} = 0,  \ 0 \leq \eta \leq n-1, \ if \ n \geq 1,
\end{eqnarray}
where $\mu_{\eta}^{(n)} \equiv \int_0^\infty d\xi  \xi^\eta \Upsilon_n(\xi)$, are the integer power moments for the $n$-th discrete state.

This power moment structure is consistent with the discrete state wavefunctions (in the $\xi \equiv \chi^2$ variable) being of the form (i.e. $b = 0$ and $\chi = x$): 
\begin{eqnarray}
\Upsilon_n(\xi) =Q_n(\xi)\Upsilon_{gr}(\xi),
\end{eqnarray}
 involving the orthonormal polynomials with respects to the weight $\Upsilon_{gr}(\chi)$; or
\begin{eqnarray}
\Psi_n(\chi) =Q_n(\chi^2)\Psi_{gr}(\chi).
\end{eqnarray}
In other words, 
\begin{eqnarray}
\int_0^\infty d\chi\ \Psi_m(\chi) \Psi_n(\chi) & = & {1\over 2}\int_0^\infty d\xi {1\over {\sqrt{\xi}}}Q_m(\xi) Q_n(\xi) \Psi_{gr}^2(\xi) \nonumber \\
&  = &  \int_0^\infty d\xi Q_m(\xi) Q_n(\xi) \Upsilon_{gr}(\xi) \nonumber \\
& = & \delta_{m,n}. \nonumber
\end{eqnarray}

That is, the orthogonality property of orthogonal polynomials in Eq.(70), relative to the weight $\Upsilon_{gr}(\xi)$, must result in the power moment behaviors given in Eq.(69).

\subsubsection{ Some  (Simple) Algebraic, EMM - $\Phi$, Bounds, $b\neq 0$}

Additional bounds on the ground state energy ensue, for arbitrary `$b$'. 

From the $p_s=0$  and $n_s = 0$ relation in Eq(61), we obtain $E_{gr}(b) >{1\over 2}$, for any $b$-value. Additionally,  if $p_s = {3\over 2}$, and $b \geq 0$, the ground state must satisfy $2E_{gr}(b)-4-2n\leq 0$, or  
$E_{gr}(b) \leq 2$. We therefore conclude, for arbitrary $ b \geq 0$ that:
\begin{equation}
{1\over 2} < E_{gr}(b) \leq    2.
\end{equation}

\subsubsection {Alternative Moment Equation Representations}

Reconsider Eq.(61) and take $p_s \equiv {\sigma\over 2}$, and define $u(n) \equiv v_{p_s}(n-2)$. We obtain
\begin{eqnarray}
 (2E-1-2(n+{\sigma\over 2})) u(n+2) = -(n+{\sigma\over2}+1/2)(n+{\sigma\over 2}-3/2)  u(n) \nonumber \\  
-2b(n+{\sigma\over 2})u(n+1),\cr
\end{eqnarray}
for $n \geq 0$. We will specifically work with $\sigma = \{0,  3\}$.  

The EMM bounding results for the ground state, based on Eq.(73), a manifestly $m_s = 1$ relation (i.e.  if $\sigma = 0$) are given in Table 2. 

By way of contrast, Eq.(73), for $\sigma = 3$, and arbitrary `$b$', is effectively an $m_s = 0$ problem since it only involves  a recursion between $\{u(1),u(2), \ldots \}$. That is, taking $n = 0$ couples $u(1)$ to $u(2)$; and taking $n=1$ then couples $u(3)$ to $u(2)$ and $u(1)$, etc. Note that the relation for $n=0$ solely pertains to the physical states. All the other moment relations ($n \geq 1$) also pertain to the unphysical states that are exponentially bounded at infinity.  This may limit its effectiveness as an eigenenergy bounding formulation (for the ground state). Indeed, this appears to be the case, as summarized in Table 2; however, despite this, the bounds for $b < O(5)$ are impressive. 

Specifically, the EMM formalism is applied to the shifted power moments: 
$\tau(n)\equiv u(n+1)$, for $n \geq 0$, resulting in:

\begin{eqnarray}
 \Big(2E-4\Big) \tau(1)  =  -3b \ \tau(0), \cr
\end{eqnarray}
and
\begin{eqnarray}
\Big(2E-4-2n\Big) \tau(n+1)  =-n(n+2)\tau(n-1) 
  -2b\big( n+{3\over 2} \big)\ \tau(n), \cr
\end{eqnarray}
for $n \geq 1$. We can then implement EMM through the $\tau(0) =1$ normalization. The results are given in Table 2. They show impressive bounds for $b < 5$.

\begin{table}
\caption{$EMM-\Phi_{gr}, m_s = 1\  (\sigma = 0), m_s = 0\  (\sigma = 3), \ from\  Eq.(73)$.}
\centerline{\begin{tabular}{rllrr}
\hline
$b$ & $E^{(L)}$    &  $E^{(U)}$  &   $P_{max}$  & Eq.[73]  \\
\hline
0 &  1.999714   &2.000244 & 27 & $\sigma =0$ \\
& 2  & 2$^*$ & 1 & $\sigma = 3$\\
\hline
.001&    1.9986710464441 & 1.9986710464498$^*$  &    16  & $\sigma = 3$\\
\hline
.01 & 1.9867452618193 & 1.9867452618204$^*$  &    20  & $\sigma = 3$\\
\hline
.1 & 1.870636 & 1.871151& 27 & $\sigma =0$ \\
 & 1.8709141846102 & 1.8709141846107$^*$ &  22 & $\sigma = 3$ \\
  \hline
.5 & 1.429056 & 1.429492 & 30 & $\sigma =0$ \\
& 1.4292927197475 & 1.4292927197522$^*$ & 24 &  $\sigma =3$\\
\hline
1 & 1.032928& 1.033250 & 27 & $\sigma =0$ \\
&1.0331033239001 &   1.0331033239766$^*$ & 25 &   $\sigma =3$\\
\hline
5 & 0.51598078 & 0.51598081 & 18 & $\sigma =0$ \\
   & 0.51591386 & 0.51598114 $^*$ &   20 & $\sigma =3$ \\
 \hline
10&0.5038074052 & 0.5038074090& 13 & $\sigma = 0$ \\
20 & 0.5009410333&  0.5009410338 & 10& $\sigma =0$ \\
100 &0.5000375056&0.5000375257& 7 & $\sigma =0$ \\
1000&0.500000375000431&0.500000375004431& 7 & $\sigma =0$ \\
2500&0.500000059999800&0.500000060003038& 6 & $\sigma=0$ \\
 \hline
$*$ & Eq.[73] with $\sigma = 3$ & is Eqs.[74,75] with  $m_s = 0$ &  & \\
\hline
\hline
\end{tabular}
}
\end{table}

\newpage

\section{EMM-$\Psi^2$ }

The previous formulations were applicable only to the ground state. An alternate strategy for obtaining bounds to the low lying discrete states is to work with the probability density, $S(\chi) \equiv \Psi^2(\chi)$. Fortunately, the probability density satisfies a third order linear ordinary differential equation (LODE). 

Given the Schrodinger equation, $-\Psi''(x) + V(x) \Psi(x) = {\cal E}\Psi(x)$, with real potential, $V(x)$, the probability density can be easily shown to satisfy 

\begin{eqnarray}
-S'''(x) + 4( V(x) - {\cal E}) S'(x)  + 2V'(x) S(x) =0.
\end{eqnarray}

In the present case, where we have $-\Psi''+({3\over 4}(x+b)^{-2}+x^2)\Psi(x) = 2E\Psi(x)$, with ${\cal E} = 2E$, we obtain:
\begin{eqnarray}
 -S'''(\chi)+\Big( {3}\chi^{-2}+4(\chi-b)^2-8E\Big) S'(\chi) + 2\Big( 2(\chi-b)-{3\over 2}\chi^{-3}  \Big) S(\chi) = 0.\cr
\end{eqnarray}

The three fundamental solutions to Eq.(77) will be (refer to Eq.(43))
\begin{eqnarray}
\Lambda_j(\chi) = \cases { Y_1^2(\chi)  = \chi^3 {\cal A}_1(\chi), \ j = 1, \cr
Y_1(\chi) Y_2(\chi) =\chi^3Ln(\chi){\cal A}_1(\chi)+\chi{\cal A}_2(\chi) , \ j = 2, \cr
Y_2^2(\chi) = \chi^3Ln^2(\chi){\cal A}_1(\chi)+2\chi Ln(\chi) {\cal A}_2(\chi)+ \chi^{-1}{\cal A}_3(\chi), \ j = 3 \cr}.\cr
\end{eqnarray}
where ${\cal A}_1(\chi) = A_1^2(\chi)$, ${\cal A}_2(\chi) = A_1(\chi) A_2(\chi)$, and ${\cal A}_3(\chi) = A_2^2(\chi)$. 

The physical solution must be of the form $\Lambda_1(\chi)$. This will have finite Stieltjes moments $v(p) = \int_0^\infty d\chi \ \chi^p \Lambda_1(\chi)$, so long as  $p >-4$. Proceeding with a similar EMM analysis to that given previously, the moment equation for the physical $S(\chi) \equiv \Psi^2(\chi)$ solutions becomes

\begin{eqnarray}
 4(1+p)v(1+p) =  
4b(1+2p)v(p)+(8E-4b^2)pv(p-1)\\ 
+(p-1)\Big(p(p-2)-3\Big)v(p-3), \nonumber \cr
\end{eqnarray}
$p > -1$. 

Let us define $v(p-3) \equiv u(p)$, it then follows that:
\begin{eqnarray}
 4(1+p)u(p+4) = 4b(1+2p)u(p+3)+(8E-4b^2)pu(p+2)\\
+(p-1)\Big(p(p-2)-3\Big)u(p), \nonumber \cr
\end{eqnarray}
$p\geq 0$.

\begin{table}
\caption{$EMM-\Psi^2, Eq.(79)$.}
\centerline{
\begin{tabular}{rrllr}
\hline
$b$ & State &$E^{(L)}$    &  $E^{(U)}$  &   $P_{max}$   \\
\hline
0 & Gr &1.99998825   &2.00005739 & 26\\
0 & $1^{st}$&3.99671544 & 4.00447362 & 25 \\
0& $2^{nd}$&5.91768300 & 6.04294214 & 26\\
\hline
.1 & Gr &1.87090202 & 1.87092049& 26 \\
.1 & $1^{st}$&3.82074273 & 3.82158281 & 28 \\
.1&$2^{nd}$&     5.75422040 & 5.82182702 & 27\\
\hline
.5 & Gr&1.42928910 & 1.42929704& 30 \\
.5& $1^{st}$&3.18388603&3.18432342&28\\
.5&$2^{nd}$&4.97456267&5.02884966&27\\
\hline
1 & Gr &          1.03310195& 1.03310458 & 29 \\
1 & $1^{st}$ &2.55658870& 2.55901222 & 27 \\
1 & $2^{nd}$&4.12209511 &4.19851998& 27 \\
\hline
5 & Gr &0.51278794 & 0.52143406 & 22 \\
5&$1^{st}$ &1.475 & 1.855 & 23 \\
 \hline
\hline
\end{tabular}}
\end{table}

The EMM-$\Psi^2$ results are given in Table 3. The results for the ground state are consistent with, although inferior to,  those in Table 2 for the EMM-$\Phi$ formulation.

\newpage

\section{Numerial Implementation of OPPQ-$\Psi$}

In Sec. 5  we outlined the generalities of the Orthonormal Polynomial Projection Quantization (OPPQ) analysis.
 Of all the three MER representations (i.e. $\Psi$, $\Phi$, $\Psi^2$), the $\Phi$ representation involved the smallest missing moment order ($m_s = 1$) as opposed to the $\Psi$ representation ($m_s = 3$). Despite this, we will implement OPPQ on $\Psi$, to avoid any confusion as to what physical configuration is being approximated.

As argued  earlier, the OPPQ-Bounding Method (OPPQ-BM)  can be used to bound, as well as approximate, the discrete state energies. An alternate approximation method, also based on the OPPQ formalism, is referred to as the Orthonormal Polynomial Projection Quantization-Approximation Method (OPPQ-AM), and is generally  easier to implement than OPPQ-BM; although the latter is the more rigorous formulation. We compare the eigenenergy approximation results of OPPQ-BM and OPPQ-AM. We also demonstrate how OPPQ-BM can be used to generate eigenenergy bounds. We only emphasize those aspects of the formalism specific to the present problem.

Although the physical wavefunction corresponds to $\Psi(\chi)$, satisfying the boundary conditions in Eq.(43), the moments in Eq.(52) are the power moments of the physical configuration ${\tilde\Psi}(\chi) \equiv \chi^{-2}\Psi(\chi)$, which behaves as $\lim_{\chi\rightarrow 0^+} {\tilde\Psi}(\chi) \sim \chi^{-{1\over 2}} A_1(\chi)$. Since the OPPQ analysis, as given below, specifically uses the MER relation for the moments of $\tilde{\Psi}$, it is this configuration that is recovered through our OPPQ analysis. 

 Consider the weighed polynomial (OPPQ) expansion for $\tilde{\Psi}$

\begin{eqnarray}
{\tilde\Psi}(\chi) = \sum_{n=0}^\infty c_n P_n(\chi) R(\chi),
\end{eqnarray}
involving the polynomials
\begin{eqnarray}
P_n(\chi) = \sum_{j=0}^n \Xi_j^{(n)} \chi^j,
\end{eqnarray}
which are required to be orthonormal with respects to an appropriate weight, $R$:
\begin{eqnarray}
\langle P_m|R|P_n\rangle = \delta_{m,n}.
\end{eqnarray}

The advantage of the weighted expansion, regardless of the choice of weight, is that the $c_n$-expansion coefficients can be generated in closed form through the associated MER relation for the power moments of $\tilde \Psi$, provided the orthonormal polynomials are known to high accuracy. For many of the classic weights, the orthonormal polynomials are known in closed form. When this is not the case, high accuracy methods based on the Cholesky decomposition of the moments of the weight are required. This will be the case for the present problem.

We recall that the MER relation for the ${\tilde \Psi}(\chi)$ physical configuration corresponds to (i.e. Eq.(55))

\begin{eqnarray}
u(p) = \sum_{\ell = 0}^{m_s = 3} M_\lambda(p,\ell) u_\ell,
\end{eqnarray}
where $\lambda \equiv 2E-b^2$, and the coefficient functions, $M_\lambda(p,\ell) \equiv M_{E,b}(p,\ell)$ are known in closed form as functions of $E$ and $b$.
  
The weighted polynomial, OPPQ, expansion coefficients are given by:
\begin{eqnarray}
c_n(E;{\overrightarrow u})  & = & \int_0^\infty d\chi  \ P_n(\chi) \ {\tilde \Psi}(\chi), \\
 & = & \sum_{j=0}^n \Xi_j^{(n)} u(j), \nonumber \\
 & = & \sum_{\ell=0}^{m_s} \Lambda_\ell^{(n)}(E) \ u_\ell, \nonumber \\
 & = & {\overrightarrow \Lambda}^{(n)}(E) \cdot {\overrightarrow u}, 
\end{eqnarray}
where
\begin{eqnarray}
\Lambda_\ell^{(n)}(E) = \sum_{j=0}^n \Xi_j^{(n)}M_E(j,\ell).
\end{eqnarray}

\subsection {An Important Observation on OPPQ's Effectiveness}

We note that one could have replaced Eq.(81) with an expansion ${\tilde \Psi}(\chi) = \sum_{n=0}^\infty d_n Q_n(\chi) R(\chi)$, utilizing the orthonormal polynomials of the square of the weight, $\langle Q_m|R^2|Q_n\rangle = \delta_{m,n}$. However, the generation of the expansion coefficients $\{d_n\}$, would require working with $\langle Q_n|R {\tilde \Psi}\rangle$, which do not offer the algebraic advantages, in general, as provided by expanding in terms of the orthonormal polynomials of $R$.  

\subsection {The OPPQ-Quantization Condition}

Quantization ensues upon demanding that the weight be chosen so that the integral 
\begin{eqnarray}
{\cal I}[{\tilde\Psi}, R] \equiv \int_0^\infty d\chi\  {{{\tilde\Psi}^2(\chi)}\over {R(\chi)}} =\cases{ finite \iff {\tilde \Psi} = {\tilde \Psi}_{phys},\cr
\infty \iff {\tilde \Psi} \neq {\tilde \Psi}_{phys},\cr}
\end{eqnarray}
 discriminate between the physical and unphysical solutions.

The natural choice for the weight is that which emulates the asymptotic form for the physical states: $R(\chi) \rightarrow {\tilde\Psi}_{phys}(\chi) \sim {\cal N}_b exp\big(-{1\over 2}(\chi-b)^2 \big)$, or more specifically:

\begin{eqnarray}
R(\chi) = \chi^{-{1\over 2}}exp\big(-{1\over 2}(\chi-b)^2 \big).
\end{eqnarray}

To generate the orthonormal polynomials, we will use a Cholesky decomposition of the Hankel moment matrix $\omega(i+j)$ constructed from the power moments of the weight $\omega(p) = \int_0^\infty d\chi \ \chi^p R(\chi)$. 

One might be tempted to use the integral identity
\begin{equation}
\int_0^\infty d\chi \chi^p R(\chi) = \Gamma(p+1/2) e^{-b^2/4}  D_{-(p+1/2)} (-b),
\end{equation}
involving the ParabolicCylinderD function, $D_\nu(z)$, to compute the power moments of the weight, and in turn,  generate the orthonormal polynomials. However, an alternate, and more efficient, strategy is to use the differential equation for the weight

\begin{equation}
\chi R'(\chi) = -{1\over 2} R(\chi) -(\chi^2-b\chi)R(\chi),
\end{equation}
in order to generate a recursive, MER, for the corresponding power moments:

\begin{equation}
\omega(p+2) = (p+{1\over 2}) \omega(p) + b \omega(p+1),
\end{equation}
$p \geq 0$. High precision calculations for $\{\omega(0),\omega(1)\}$ generate all the other power moments.

The quantization integral condition in Eq.(88) can be easily transformed into a positive series expansion: 

\begin{eqnarray}
 {\cal I}[{\tilde\Psi}, R] & = & \sum_{n=0}^\infty \ c_n^2 [E,{\overrightarrow u}] \\
& = & \cases{finite, \iff \ E= E_{phys}\ and \ u_\ell = u_{\ell,phys}, \cr
			\infty, \iff \ E \neq E_{phys}\ or \ u_\ell \neq u_{\ell,phys} \cr}.
\end{eqnarray}

\subsection {The OPPQ-Approximation Method}
If the series in Eq.(93) is to be finite for the exact physical energy and associated missing moments, then it follows that the coefficients, for the physical parameter values, must satisfy:

\begin{eqnarray}
Lim_{n\rightarrow \infty} c_n[E_{phys},{\overrightarrow u}_{phys}] = 0.
\end{eqnarray}
We can use this to approximate the physical parameters by taking $n = N-\ell$, for (large) fixed $N$ and $0 \leq \ell \leq m_s \ (i.e. \ 3) $. Using Eq.(86) we obtain
\begin{eqnarray}
c_{N-\ell_1}(E; {\overrightarrow u}) = {\overrightarrow \Lambda}^{(N-\ell_1)}(E)\cdot {\overrightarrow u} = 0,
\end{eqnarray}
resulting in the determinantal condition:
\begin{eqnarray}
Det\Big( \Lambda_{\ell_2}^{(N-\ell_1)}(E) \Big) = 0,
\end{eqnarray}
where $0 \leq \ell_{1,2} \leq 3$. This will yield approximants for the energy, to  the $N$-th expansion order. This defines the OPPQ-Approximation Method, or OPPQ-AM. Note, that the, approximate, physical missing moments are assumed to be the (only) null vector for the $(1+m_s)\times(1+m_s)$ matrix,  $ \Lambda_{\ell_2}^{(N-\ell_1)}(E)$, for $E$ values satisfying Eq.(97).

\subsubsection{Numerical OPPQ-AM Results}

In Table 4 we show the results of implementing OPPQ-AM for various values of $b: 0 \rightarrow 10$. The results in Table 4 are stable (i.e. convergng) to many more digits than given; and correspond to using moment expansion order $P_{max} = 100$.  We note the exact accuracy of the energies for the $b = 0$ case, as explained below. Comparing the $E_0$ entries for $b = .5,1, 5, 10$ with those in Table 2 (derived through EMM), shows that OPPQ-AM results are consistent with the tight EMM bounds. The same applies for the excited states, with bounds reported in Table 3, also derived through EMM-$\Psi^2$. The results in Table 4 are graphically illustrated in Fig. 1. These results strongly suggest that in the $b\rightarrow \infty$ limit, we recover the pure harmonic oscillator states with energies (based on our choice of notation) $E_n(b=\infty) = (n+{1\over 2})$.

\begin{table}
\caption{
$OPPQ-AM, P_{max} = 100$.
}
\centerline{
\begin{tabular}{rrrrr}
\hline
$b$ & $E_0$    &  $E_1$  &   $E_2$  & $E_3$\\
\hline
0& 2 & 4  & 6 & 8 \\
.5&  1.4292927197    &    3.184017114	& 4.987971463 &	6.820440707\\
1.0&1.0331033239	    & 2.557261915	&4.169923329	&  5.837014390\\
1.5&0.7847675572   &2.107433725	&3.538491138	&5.044354682\\
2.0&0.6481322228  &	1.816590914	 &3.084658976	&4.436894490\\
2.5&0.5818553905	&1.655297046	&2.794166923	&4.007820744\\
3.0&0.5509509520	   &1.580121756	&2.638483895	&3.743614149\\
3.5&0.5351717068	&1.547741639	&2.570043163	&3.611459829\\
4.0&0.5259688826	&1.532318972	&2.541876785	&3.557419442\\
4.5&0.5200471427	&1.523670892	&2.528557218	&3.535449712\\
5.0&0.5159807819	&1.518222436	&2.521046746	&3.524694536\\
5.5&0.5130560157	&1.514525084	&2.516293284	&3.518454199\\
6.0&0.5108767399	&1.511882954	&2.513054951	&3.514433855\\
6.5&0.5092067728	&1.509920644	&2.510731970	&3.511660509\\
7.0&0.5078974472	&1.508418730	&2.509000073	&3.509651653\\
7.5&0.5068510767	&1.507241028	&2.507669464	&3.508141930\\
8.0&0.5060011798	&1.506298942	&2.506622186	&3.506974079\\
8.5&0.5053011643	&1.505532597	&2.505781384	&3.506049402\\
9.0&0.5047175513	&1.504900234	&2.505095024	&3.505303077\\
9.5&0.5042257611	&1.504371940	&2.504526749	&3.504690917\\
10.0&0.5038074053	&1.503925798	&2.504050459	&3.504181861\\
 \hline
\hline
\end{tabular}}
\end{table}
\newpage
\begin{figure}
\includegraphics{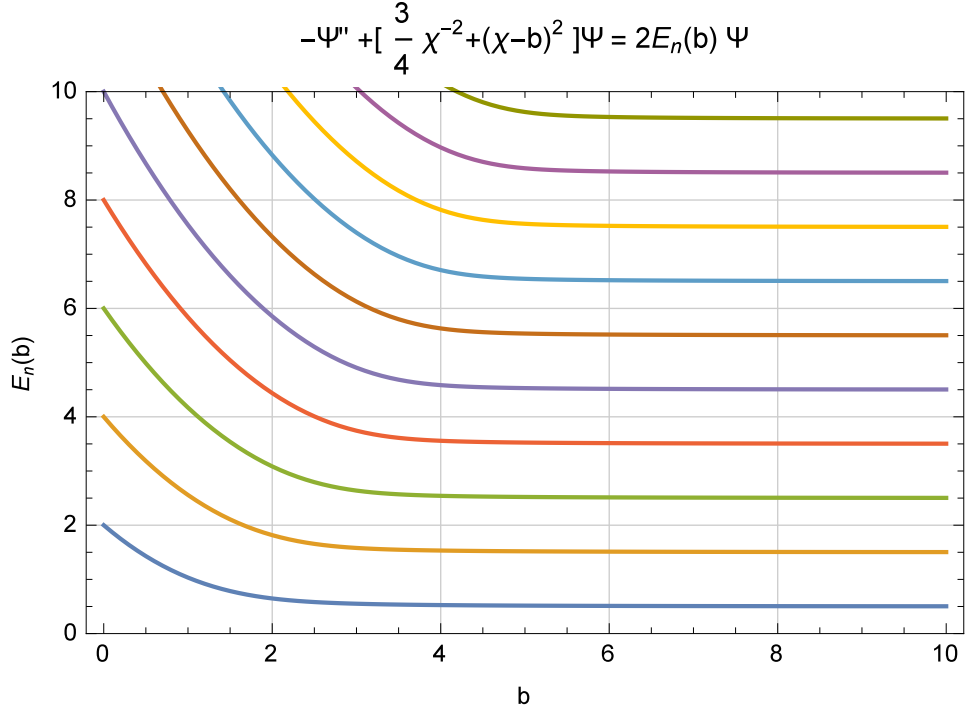}
\centering{\caption{Discrete state energy plots, $E_n(b)$, for $0 \leq n \leq 9$ and $0 \leq b \leq 10$.}}
\end{figure}

\subsubsection{OPPQ Confirmation of the Existence of Exactly Solvable Solutions for $b = 0$}

As argued earlier, the $b = 0$ case corresponds to an exactly solvable system  in which the true physical solutions are of the form ${\Psi}_n(\chi) = Q_n(\chi^2) { \Psi}_0(\chi)$ (i.e. Eq.(71)) involving (a subset of) the orthonormal polynomials relative to the square of the weight, ${\Psi}_0^2(\chi)$ (i.e. $\langle Q_m|{\Psi}_0^2|Q_n\rangle = \delta_{m,n}$) as defined relative to the variable $\chi$. That is, the $\{ Q_n(\xi)\}$ are all the orthonormal polynomials of the weight ${1\over {2\sqrt{\xi}}}\Psi_{gr}^2(\xi)$, where $\xi \equiv \chi^2$. 

We then have ${\tilde \Psi}_n(\chi) \equiv \chi^{-2}\Psi_n(\chi) = Q_n(\chi^2) R(\chi)$, where $R(\chi) = \chi^{-2} \Psi_{gr}(\chi)$ (i.e. Eq.(89)). For these exact configurations, the OPPQ expansion truncates, ${\tilde \Psi}_n(\chi) = \sum_{\eta = 0}^{2n} c_{\eta}^{(n)} P_\eta(\chi) \ R(\chi)$, involving the orthonormal polynomials of $R(\chi)$.   Therefore, for a given, {\it exactly solvable}, discrete state energy, $E_{phys;n}$, we would expect  Eq.(97) to behave as
\begin{eqnarray}
c_{N-\ell_1}(E_{phys;n}; {\overrightarrow u}) = {\overrightarrow \Lambda}^{(N-\ell_1)}(E_{phys;n})\cdot {\overrightarrow u} = 0,
\end{eqnarray}
for $N-(m_s+1) \geq 2n+1$. For $b \neq 0$, the MER relation in Eq.(52) is of order $m_s = 3$; however, for $b = 0$ it effectively becomes $m_s = 1$. The results in Table 5 are in keeping with $N \geq 2n+4$. That is, the OPPQ-AM determinant must be a  polynomial with roots at $E = E_n$, for $N \geq 2n+4$.

\begin{table}
\caption{
$Det\Big( \Lambda_{\ell_2}^{(N-\ell_1)}(E) \Big)$, for $b = 0$.
}
\centerline{
\begin{tabular}{cc}
\hline
$N$ & $Det^{(N)}(E) \propto$ \\
\hline
4 &  $(E-2) $\\
5&   $(E-2)(E-4.254) $   \\
6&   $(E-2)(E-4)(E-7.463) $   \\
7&   $(E-2)(E-4)(E-5.708)(E-12.632) $   \\
8&   $\Pi_{n=0}^2(E-2(n+1))\times Poly(degree\ 2)$\\  
10& $\Pi_{n=0}^3(E-2(n+1))\times Poly(degree\ 3)$\\  
20& $\Pi_{n=0}^8(E-2(n+1))\times Poly(degree\ 8)$\\
30&$\Pi_{n=0}^{13}(E-2(n+1))\times Poly(degree\ 13)$\\
\hline
\hline
\end{tabular}}
\end{table}

\subsection{Implementing the OPPQ-Bounding Method (BM)}

As explained in Sec. 5, the focus of the OPPQ-BM formalism, for one dimensional systems, are the lowest eigenvalues of the positive definite matrices 

\begin{eqnarray}
{\cal P}^{(n)}(E) = \sum_{\eta=0}^n {\overrightarrow \Lambda}^{(\eta)}(E) {\overrightarrow \Lambda}^{(\eta)}(E),
\end{eqnarray}
where
\begin{eqnarray}
\lambda_n(E) = \ Smallest\ Eigenvalue \ of \ {\cal P}^{(n)}(E).
\end{eqnarray}
These expressions form a positively increasing sequence, for any $E$:
\begin{eqnarray}
0 < \lambda_n(E) < \lambda_{n+1}(E) < \lambda_{n+2}(E) < \ldots \lambda_\infty(E).
\end{eqnarray}
Thus, the eigenvalue functions, $\lambda_n(E)$, form a nested, concaved (upwards), sequence of functions that become unbounded everywhere except at the physical energies. We show the progression of the $\lambda_n(E)$ nested function sequence in Figs. 2-5, for the $b=.5$ case, and the first four energy levels. 

 In Fig. 2 we show the nesting (i.e. monotonically increasing) nature of the $\lambda_n(E)$ functions for $10 \leq n \leq 20$. It is important to recognize that the extrema locations in the energy parameter variable (i.e. $\partial_E\lambda_n(E_n^{(min)}) = 0$), do not necessarily behave in a monotic manner until (perhaps) they converge to the physical discrete energy value. However, the value of $\lambda_n(E_n^{(min)})$ do converge, from below, monotonically to the correct physical value, in the infinite expansion limit. This is best appreciated from Table 6, which captures all the extrema information in Figures 2-5.

\begin{figure}
\center{\includegraphics{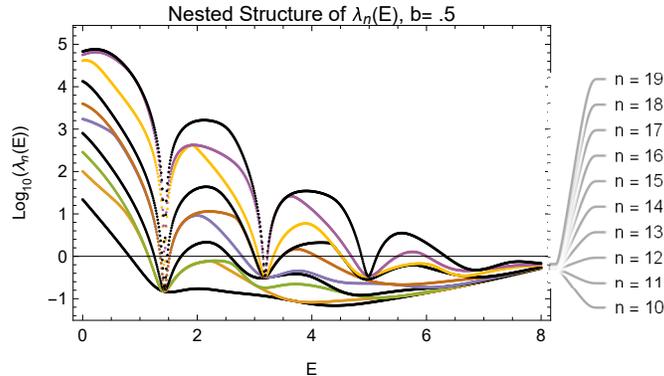}}
\centering{\caption{Nested $\lambda_n(E)$ sequence, $n = 10, 11, \ldots, 19$, $0 \leq E \leq 8$, and $b = .5$ (i.e. Eq.(101))}}
\end{figure}

\begin{figure}
\center{\includegraphics{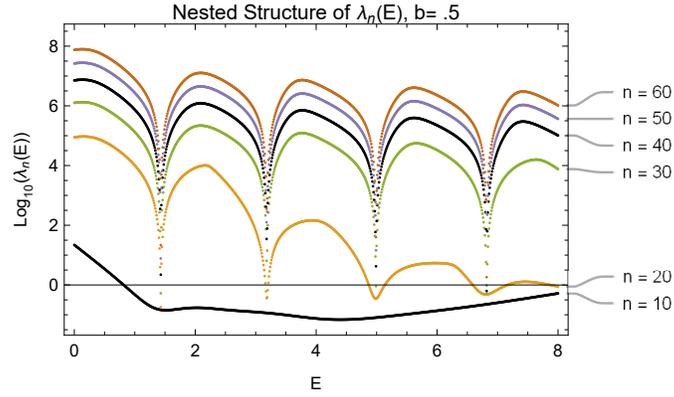}}
\centering{\caption{Nested $\lambda_n(E)$ sequence, $n = 10, 20, \ldots, 60$, $0 \leq E \leq 8$, and $b = .5$ (i.e. Eq.(101))}}
\end{figure}

\begin{figure}
\center{\includegraphics{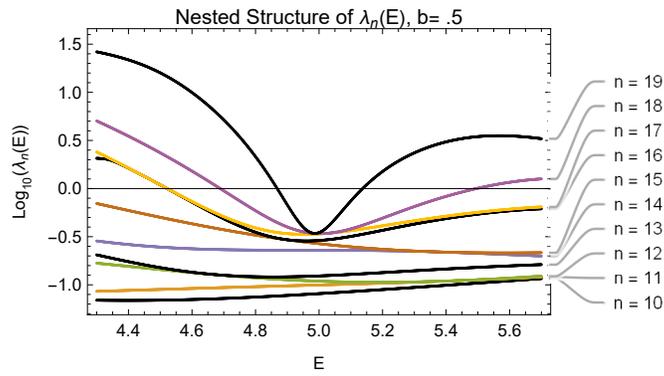}}
\centering{\caption{Nested $\lambda_n(E)$ sequence, $n = 10, 11, \ldots, 19$, $4.4 \leq E \leq 5.6$, and $b = .5$ (i.e. Eq.(101))}}
\end{figure}

\begin{figure}
\center{\includegraphics{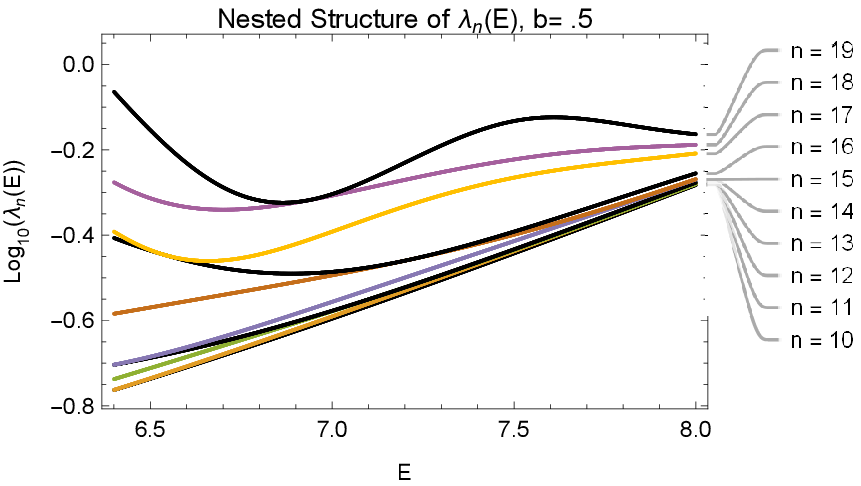}}
\centering{\caption{Nested $\lambda_n(E)$ sequence, $n = 10, 11, \ldots, 19$, $6.5 \leq E \leq 8.0$, and $b = .5$ (i.e. Eq.(101))}}
\end{figure}

\begin{sidewaystable}
\caption{
OPPQ-BM Generated Energy Approximants: $\partial_E\lambda_N(E_N^{(min)}) = 0$, and ${\cal L}_{10}(E_N^{(min)}) \equiv Log_{10}\big(\lambda_N(E_N^{(min)})\big)$, for $ b = .5$
}
\small
\resizebox{\columnwidth}{!}
\centerline{
\begin{tabular}{rllllllll}
\hline
$N$& $E_{N;0}^{(min)}$ & ${\cal L}_{10}(E_{N;0}^{(min)}) $ &$ E_{N;1}^{(min)}$ &${\cal L}_{10}(E_{N;1}^{(min)}) $ & $E_{N;2}^{(min)}$ &${\cal L}_{10}(E_{N;2}^{(min)}) $&$ E_{N;3}^{(min)}$ & ${\cal L}_{10}(E_{N;3}^{(min)}) $ \\
\hline
10&  1.5150470& -0.84559280  & 4.3969969 & -1.1623635        &     &    &      &        \\
11&  1.4199646 & -0.82806681&  3.9889962  &  -1.0852239       &   &       &    &     \\
12 & 1.4156228 & -0.80336151&  3.1875626 &  -0.74234366 & 5.1889348 & -0.97331160&  & \\
13 & 1.4301144  & -0.79832277&  3.1214643 & -0.51419926 & 4.8564978 & -0.91784021&  &\\
14 & 1.4290630 &  -0.79825893& 3.2393572 & -0.49476172 &6.0745714 &   -0.73221637&  & \\
15 & 1.4289479 &  -0.79757698& 3.1808773 & -0.48495873 & 5.5858837 &  -0.66854618&  & \\
16 & 1.4294188&  -0.79745638&  3.1790069 &-0.47258623 & 4.9600850 &  -0.54416892 & 6.8883817  & -0.49025962\\
17 & 1.4293205 &  -0.79744519& 3.1863572 &  -0.47115027&4.9543391 &-0.47566983&  6.6591851 & -0.46057027\\
18 & 1.4292787 &  -0.79740046&  3.1841563  & -0.47087295 & 5.0040638 & -0.46768132&6.7008468 & -0.34041178\\
19 & 1.4292932 & -0.79738417&   3.1837027&  -0.47025848&4.9868242  & -0.46552191& 6.8641735 & -0.32407160\\
20 & 1.4292967&  -0.79738248 &  3.1840333 & -0.47013608 & 4.9851930  & -0.46121438&  6.8069551 &-0.31888442\\
\hline
30&  1.4292931 &-0.79738040& 3.1840182 &-0.47012365& 4.9879738&-0.46026542&6.8204428&-0.30655075\\
40& 1.4292928& -0.79738016&3.1840173 &-0.47012319 &4.9879718&-0.46026461& 6.8204411 &-0.30655027\\
50& 1.4292927 & -0.79738012&3.1840172 &  -0.47012312&4.9879716&-0.46026449&6.8204408&-0.30655019\\
60& 1.4292927&-0.79738011&3.1840171&-0.47012310&4.9879715 &-0.46026446&6.8204408&-0.30655017\\
70 & 1.42929272246 &-0.79738011&  3.18401712169&  -0.47012309& 4.98797147862& -0.46026445&6.82044072541 & -0.30655016\\
80 & 1.42929272103 &-0.79738011&  3.18401711764&  -0.47012309& 4.98797147036& -0.46026444&6.82044071591 & -0.30655016\\
90 & 1.42929272042 & -0.79738011&3.18401711589& -0.47012309&    4.98797146679& -0.46026444&6.82044071179& -0.30655016\\
100&1.42929272012&  -0.79738011&3.18401711506 &-0.47012309&    4.98797146508& -0.46026444&6.82044070983& -0.30655016\\
\hline
150&1.42929271979&  -0.79738011&3.18401711412 &-0.47012309&    4.98797146316& -0.46026444&6.82044070761& -0.30655016\\
200&1.42929271976&  -0.79738011&3.18401711403 &-0.47012309&    4.98797146298& -0.46026444&6.82044070740& -0.30655016\\
250&1.429292719754&  -0.79738011&3.18401711401 &-0.47012309&    4.98797146294& -0.46026444&6.82044070736&-0.30655016\\
300&1.429292719752&  -0.79738011&3.18401711401 &-0.47012309&    4.98797146293& -0.46026444&6.82044070735&-0.30655016\\
350&1.4292927197517&  -0.79738011&3.18401711400 &-0.47012309&    4.98797146293& -0.46026444&6.82044070734&-0.30655016\\
\hline
 \hline
${\cal B}_U$ &    & $10^{-.79738}$ &  & $10^{-.470123}$ &  &$10^{ -.460264}$&  &$10^{-.30655}$\\
\hline
\end{tabular}}
\end{sidewaystable}

\subsubsection{Using OPPQ-BM To Approximate the Eigenenergies} 

As previously noted, the local minima
\begin{eqnarray}
\partial_E\lambda_n(E_n^{(min)}) = 0,
\end{eqnarray}
define the $n$-th order (OPPQ-BM) approximant to the discrete state energy. These extrema are necessarily real. Also, one does not have to numerically determine these derivatives. It is straightforward to generate an algebraic procedure for generating the function $\partial_E\lambda_n(E)$, and then determine its zeroes, and in particular the local minima. The results of this analysis (as given in Table 6) concur with the data in Table 4. 

More specifically, the results in Table 4, based on OPPQ-AM, were generated based on a maximum moment expansion order of $P_{max} = 100$. With regards to the ground state energy, for $b = .5$, we see that the OPPQ-AM results in Table 4 ($E_{gr}(b=.5) = 1.4292927197$)  concur with the EMM bounds in Table 2 ($1.4292927197475 < E_{gr} < 1.4292927197522$) generated on the basis of maximum moment expansion order $P_{max} = 24$ (based on Eq.[73] with $\sigma = 3$). However, within the OPPQ-BM's Approximation Ansatz (i.e. Eq. (102)), it takes $P_{max} = 150$ (i.e. Table 6) to achieve comparable results to OPPQ-AM for $P_{max} = 100$. It takes OPPQ-BM's Approximation Ansatz a $P_{max} = 350$ to compete with the EMM bounds in Table 2 (obtained with $P_{max} = 24$). Despite all this, EMM-$\Psi^2$ cannot give the same tightness of bounds for the excited states, as  demonstrated in Table 3. Fortunately, through OPPQ-BM, we can achieve vastly improved (tighter) bounds. Indeed, the OPPQ-BM bounds for the discrete states at $b = .5$, as given in Table 7, tightly bound the OPPQ-AM results in Table 4. This is discussed below.
\\
\\

\subsubsection{OPPQ-BM: Generating Eigenenergy Bounds}

The local minima in Eq.(102) also serve to define an increasing positive sequence that is bounded from above by the true physical energy counterpart:

\begin{eqnarray}
\lambda_n(E_n^{(min}) < \lambda_{n+1}(E_{n+1}^{(min})  < \ldots < \lambda_\infty(E_{phys}) < \infty.
\end{eqnarray}
Let ${\cal B}_U$ be any, empirically determined, coarse upper bound to the  convergent sequence in Eq.(103) (i.e. for a chosen discrete state):
\begin{eqnarray}
\lambda_n(E_n^{(min}) < \lambda_{n+1}(E_{n+1}^{(min})  < \ldots < \lambda_\infty(E_{phys}) < {\cal B}_U.
\end{eqnarray}
As argued in Sec. 5, there will always be $E$ values satisfying:
\begin{eqnarray}
\lambda_n(E_n^{(L)}) = \lambda_n(E_n^{(U}) = {\cal B}_U.
\end{eqnarray}
These will then correspond to converging lower and upper bounds to the  desired physical energy:
\begin{eqnarray}
E_n^{(L)} < E_{phys} < E_n^{(U)},
\end{eqnarray}
with
\begin{eqnarray}
Lim_{n \rightarrow \infty} \Big( E_n^{(U)}-E_n^{(L)} \Big) = 0^+.
\end{eqnarray}

The results of this bounding analysis on the  first four discrete states for $b = .5$, utilizing the coarse upper bounds ${\cal B}_U$ appearing at the bottom of Table 6, are given in Table7. We note that these are far superior to those in Table 3, based on EMM-$\Psi^2$. Keeping these coarse ${\cal B}_U$ upper bounds, we can generate arbitrarily converging lower and upper bounds, as $n \rightarrow \infty$.

\begin{sidewaystable}
\caption{
OPPQ-BM Upper and Lower Bounds: $E_N^{(L)} < E_{phys} < E_N^{(U)}$,  for $ b = .5$
}
\small
\resizebox{\columnwidth}{!}
\centerline{
\begin{tabular}{rllllllll}
\hline
$N$& $E_{0}^{(L)}$ & $E_{0}^{(U)}$ &$E_{1}^{(L)}$ & $E_{1}^{(U)}$  & $E_{2}^{(L)}$ & $E_{2}^{(U)}$&$E_{3}^{(L)}$ & $E_{3}^{(U)}$ \\
\hline
10&   1.355213912&     1.767314750  &  &        &     &    &      &        \\
50&   1.429292680  &   1.429292800&     3.184017055   &  3.184017276      &   4.987971200 &     4.987972000     & 6.820440400   & 6.820441200  \\
100& 1.4292927126 &  1.4292927276&   3.184017100   &  3.184017130 &        4.987971416  &    4.987971512&      6.820440664  &  6.820440752 \\
150& 1.4292927172 &  1.4292927224&   3.1840171096 &  3.1840171186 &       4.9879714476 &   4.9879714784&   6.8204406948  & 6.8204407212\\
\hline
 \hline
\end{tabular}}
\end{sidewaystable}

\subsubsection {Wavefunction Reconstruction}

The OPPQ expansion of the wavefunction $\Psi(\chi) \equiv \chi^2 {\tilde \Psi}(\chi)$ is obtained from Eq.(81) and Eq.(89). The wavefunction reconstruction uses all the previously generated results (i.e. physical energy approximants, etc) but only uses the OPPQ-expansion coefficients, $\{c_n| n \leq 40\}$. We plot the results for the first four discrete state energy levels corresponding to $b=\{0,1,2,3,4,5\}$ in order to visualize the evolution of the wavefunctions relative to the anticipated harmonic osciallator solutions as $b \rightarrow \infty$. All of this is depicted in Figures 6-11. It will be noticed that relative to the point $\chi=b$ (i.e. $x = 0$) , the wavefunctions become more symmetric, or antisymmetric, as $b \rightarrow 5$.

\begin{figure}
\center{\includegraphics{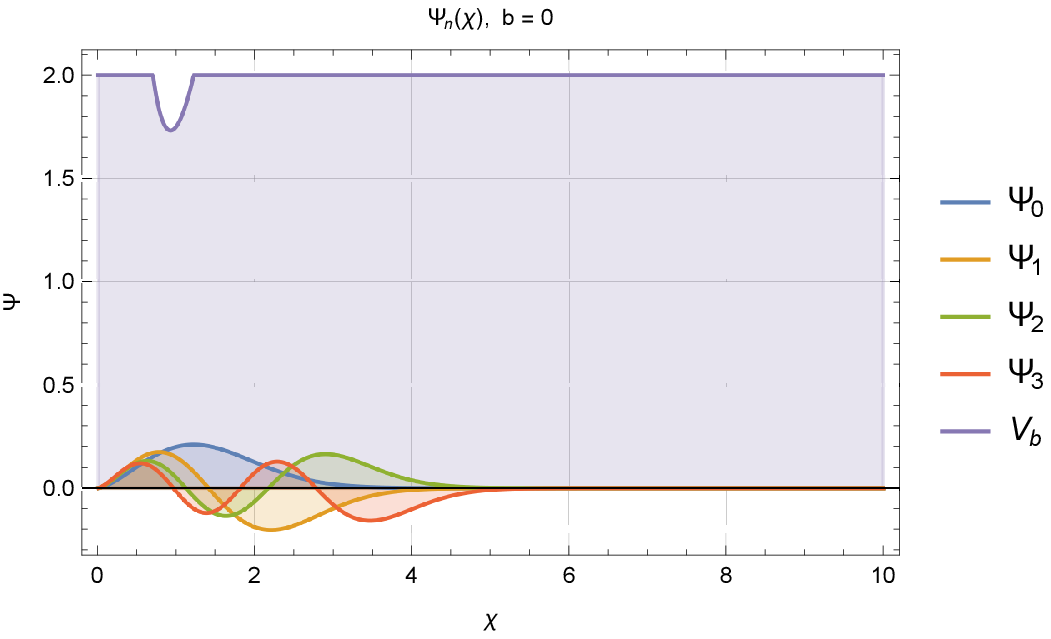}}
\centering{\caption{ Discrete state wavefunctions, $\Psi_n(\chi)$, for $n \in \{0,1,2,3\}$; and potential, $V_b(\chi) = {1\over 2}\Big({3\over 4}\chi^{-2} + (\chi-b)^2\Big)$, for $b = 0$}}
\end{figure}

\begin{figure}
\center{\includegraphics{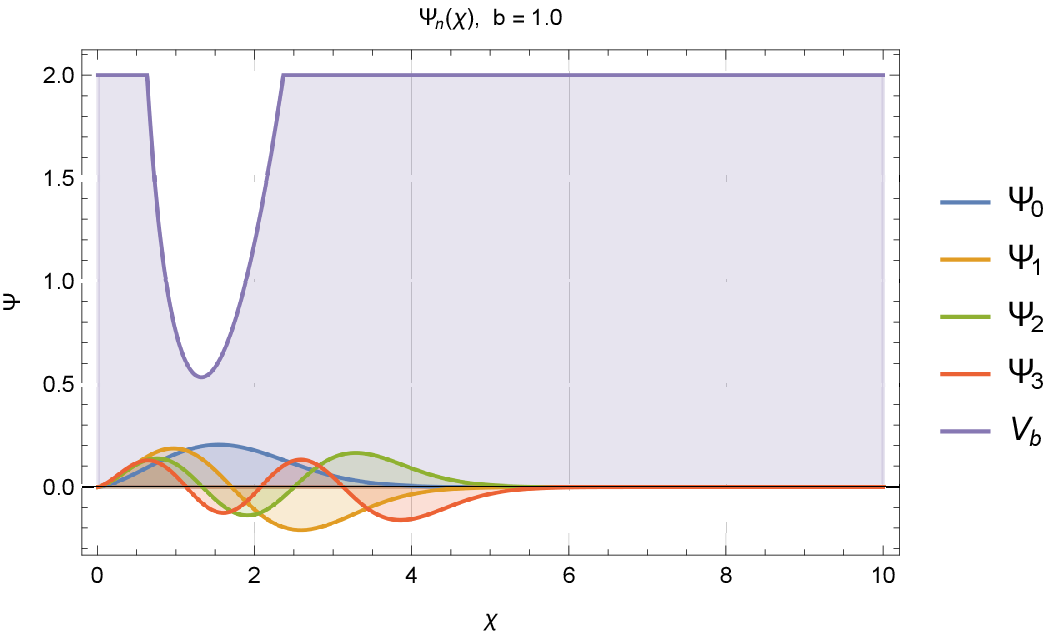}}
\centering{\caption{ Discrete state wavefunctions, $\Psi_n(\chi)$, for $n \in \{0,1,2,3\}$; and potential, $V_b(\chi) = {1\over 2}\Big({3\over 4}\chi^{-2} + (\chi-b)^2\Big)$, for $b = 1$}}
\end{figure}

\begin{figure}
\center{\includegraphics{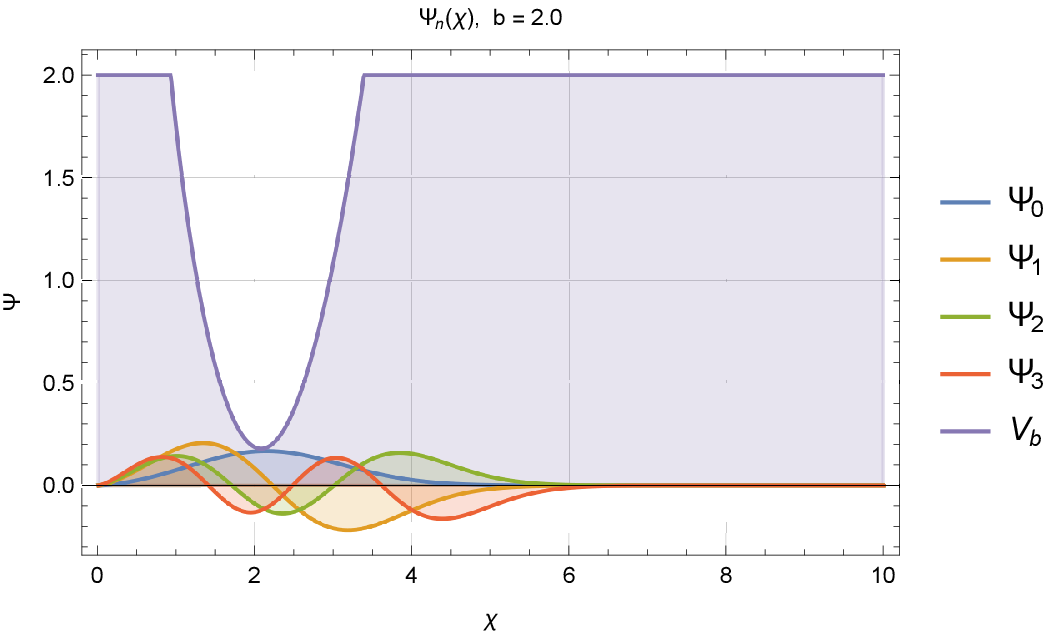}}
\centering{\caption{ Discrete state wavefunctions, $\Psi_n(\chi)$, for $n \in \{0,1,2,3\}$; and potential, $V_b(\chi) = {1\over 2}\Big({3\over 4}\chi^{-2} + (\chi-b)^2\Big)$, for $b = 2$}}
\end{figure}

\begin{figure}
\center{\includegraphics{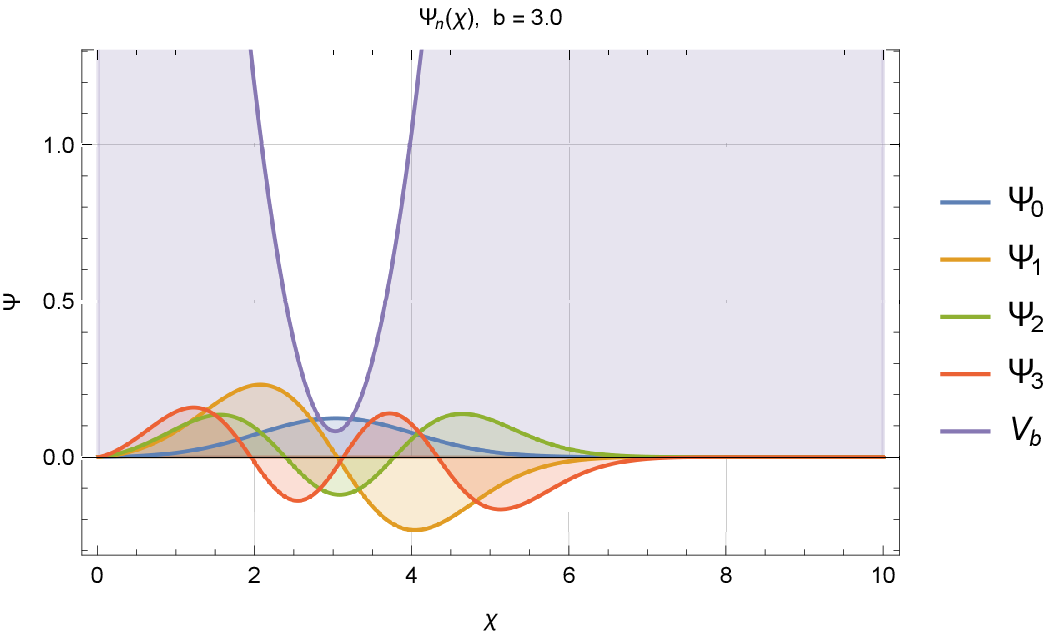}}
\centering{\caption{ Discrete state wavefunctions, $\Psi_n(\chi)$, for $n \in \{0,1,2,3\}$; and potential, $V_b(\chi) = {1\over 2}\Big({3\over 4}\chi^{-2} + (\chi-b)^2\Big)$, for $b = 3$}}
\end{figure}

\begin{figure}
\center{\includegraphics{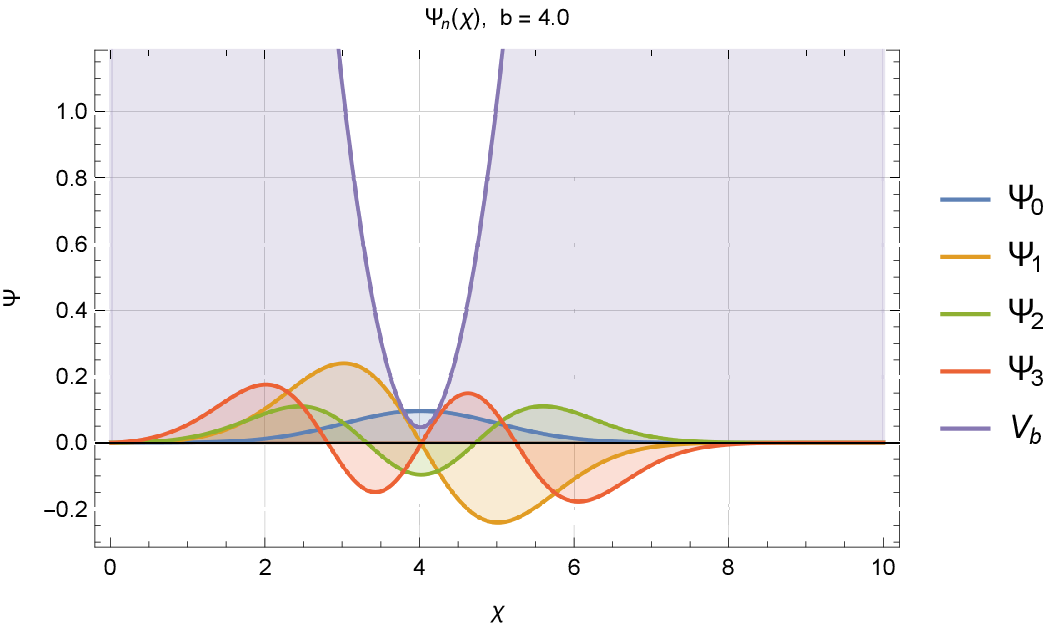}}
\centering{\caption{ Discrete state wavefunctions, $\Psi_n(\chi)$, for $n \in \{0,1,2,3\}$; and potential, $V_b(\chi) = {1\over 2}\Big({3\over 4}\chi^{-2} + (\chi-b)^2\Big)$, for $b = 4$}}
\end{figure}

\begin{figure}
\center{\includegraphics{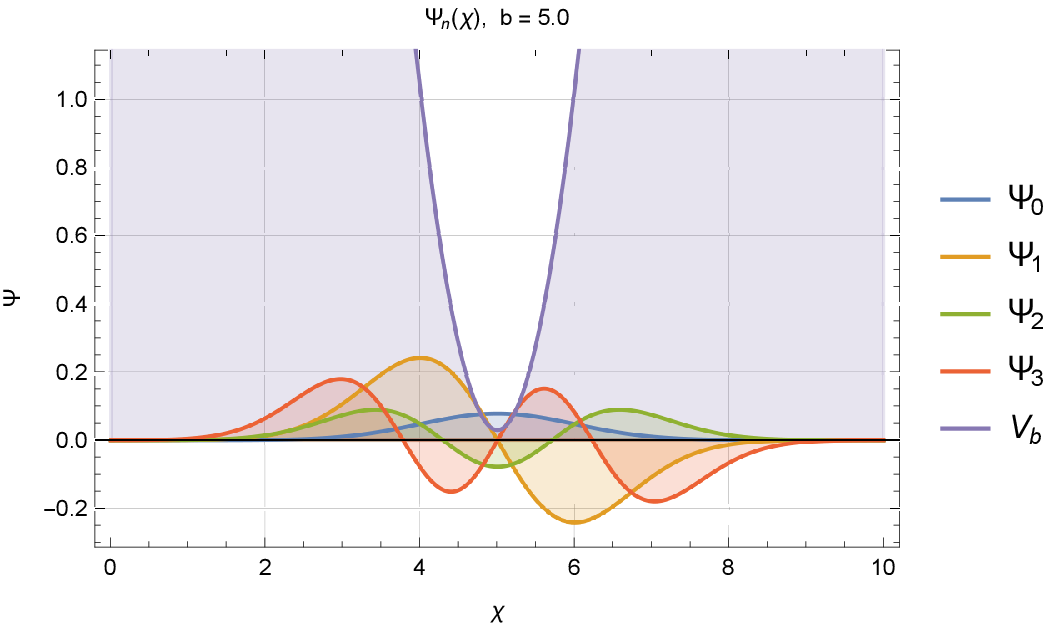}}
\centering{\caption{ Discrete state wavefunctions, $\Psi_n(\chi)$, for $n \in \{0,1,2,3\}$; and potential, $V_b(\chi) = {1\over 2}\Big({3\over 4}\chi^{-2} + (\chi-b)^2\Big)$, for $b = 5$}}
\end{figure}



\newpage

\section{References}
\noindent [1] Gouba L, 2021 {\em Journal of High Energy Physics, Gravitation and Cosmology}
 {\bf 7} 352-365; DOI:10.4236/jhepgc.2021.71019\\

\noindent [2] Nikiforov A F and Uvarov V B 1988 {\em{Special functions of Mathematical Physics}} (Birkhauser, Boston)\\

\noindent [3] Mhaskar H N 1996 {\em{Introduction to the theory of Weighted Polynomial Approximation}} {(Singapore: World Scientific Pub. Co. Inc)}\\

\noindent [4] Handy C R and Vrinceanu D 2013 {\em J. Phys. A: Math. Theor. } {\bf 46} 135202 \\

\noindent [5] Handy C R and Vrinceanu D 2013 {\em J. Phys. B: At. Mol. Opt. Phys.} {\bf 46} 115002 \\ 

\noindent [6] Handy C R 2021 {\em Phys. Scr.} {\bf 96} 075201\\

\noindent [7] Handy C R and Bessis D 1985 {\em Phys. Rev. Lett.} {\bf 55} 931 \\

\noindent [8] Bessis D and Handy C R 1986 { \em Int. J. of Quantum Chem: Quantum Chem. Symp.} {\bf 20}, 21 (John Wiley \& Sons, Inc.)\\

\noindent [9] Handy C R, Bessis D, Sigismondi G, and Morley T D 1988 {\em Phys. Rev. A} {\bf 37} 4557 \\

\noindent [10] Handy C R, Bessis D, Sigismondi G, and Morley T D 1988 {\em Phys. Rev. Lett.} {\bf 60} 253 \\

\noindent [11] Boyd S and Vandenberghe L 2004 {\it Convex Optimization} (New York: Cambridge University Press)\\

\noindent [12] Lasserre J-B 2009 {\it { Moments, Positive Polynomials and Their Applications}} (London: Imperial College Press )\\

\noindent [13] Chvatal V 1983 {\it Linear Programming} (Freeman, New York)\\

\noindent [14] Shohat J A and Tamarkin J D, 1963 {\it The Problem of Moments} (American Mathematical Society, Providence, RI) \\

\noindent [15] Handy C R 2001 {\em J. Phys. A} {\bf 34} 5065\\

\noindent [16]Handy C R and Vrinceanu D 2016 {\em Can. J. of Physics}, 94(4): 410-424.\\

\noindent [17] Handy C R, Vrinceanu D, and  Gupta R, 2014 {\em J. Phys. A: Math.Theor. } {\bf 47}: 295203\\

\noindent [18].Handy C R, Msezane A Z, and  Yan Z, 2002 {\em J. Phys. A: Math. Gen.} {\bf 35}, 6359  \\

\noindent [19]. Handy C R, Tymczak C J, and Msezane A Z, 2002 {\em Phys. Rev. A} {\bf 66}, 050701 (R)\\

\noindent [20] Koosis P 1988 {\em{\it The Logarithmic Integral I}} { Chapter VI, pg 170} (Cambridge Studies in Adv. Math., Cambridge Univ. Press)\\

\end{document}